\documentclass[10pt,final,journal,letterpaper,twoside,twocolumn]{IEEEtran}

\newenvironment{proofwl}[1]{\noindent\hspace{2em}{\itshape #1: }}%
{\hspace*{\fill}~\QED\par\endtrivlist\unskip}

\usepackage{amsmath}
\usepackage{amssymb}

\ifx\pdfoutput\undefined
 \usepackage[dvips]{graphicx}
\else
 \usepackage[pdftex]{graphicx}
\fi

\usepackage{cite}
\usepackage{enumerate}

\newtheorem{lause}{Theorem}
\newtheorem{seur}{Corollary}
\newtheorem{lemma}{Lemma}
\newtheorem{esim}{Example}
\newtheorem{maar}{Definition}
\newtheorem{huom}{Remark}

\newcommand{\realpart}[1]{{#1}_R}
\newcommand{\imagpart}[1]{{#1}_I}

\newcommand{\randomtype}[1]{\mathtt{#1}}
\newcommand{\sourcecomp}{\randomtype{s}}
\newcommand{\source}{\vec{\sourcecomp}}
\newcommand{\altsourcecomp}{\randomtype{r}}
\newcommand{\altsource}{\vec{\altsourcecomp}}
\newcommand{\mixturecomp}{\randomtype{x}}
\newcommand{\mixture}{\vec{\mixturecomp}}
\newcommand{\solucomp}{\randomtype{y}}
\newcommand{\soluvec}{\vec{\solucomp}}
\newcommand{\realrvc}{\realpart{\mixture}}
\newcommand{\imagrvc}{\imagpart{\mixture}}
\newcommand{\normal}{\randomtype{n}}
\newcommand{\mnormal}{\vec{\normal}}
\newcommand{\snormal}{\randomtype{\eta}}
\newcommand{\smnormal}{\vec{\snormal}}

\newcommand{\compvector}{{\mathbf z}}
\newcommand{\cocomp}{z}
\newcommand{\recomp}{t}
\newcommand{\cconst}{c}
\newcommand{\cconsttwo}{d}
\newcommand{\cconstthree}{a}
\newcommand{\cconstfour}{b}
\newcommand{\rconst}{\alpha}
\newcommand{\cargu}{\theta}
\newcommand{\realvz}{\realpart{\compvector}}
\newcommand{\realz}{\realpart{\cocomp}}
\newcommand{\imagvz}{\imagpart{\compvector}}
\newcommand{\imagz}{\imagpart{\cocomp}}
\newcommand{\isoreal}[1]{{#1}_{\mathbb{R}}}

\newcommand{\dindex}{k}
\newcommand{\dindextwo}{l}
\newcommand{\dindexthree}{q}

\newcommand{\poly}{{\mathcal P}}

\newcommand{\pinv}{\#}

\newcommand{\cconj}[1]{#1^*}
\newcommand{\inner}[2]{\langle #1,#2\rangle}
\newcommand{\norm}[1]{\parallel\!\!#1\!\!\parallel}
\newcommand{\htrans}{H}
\newcommand{\trans}{T}
\DeclareMathOperator{\Reop}{Re}
\newcommand{\real}[1]{\Reop\bigl\{#1\bigr\}}
\DeclareMathOperator{\Imop}{Im}
\newcommand{\imag}[1]{\Imop\bigl\{#1\bigr\}}
\DeclareMathOperator{\Eop}{E}
\newcommand{\E}[2]{\Eop_{#1}\bigl[#2\bigr]}
\DeclareMathOperator{\covop}{cov}
\newcommand{\twocov}[2]{\covop\bigl[#1,#2\bigr]}
\newcommand{\cov}[1]{\covop\bigl[#1\bigr]}
\DeclareMathOperator{\pcovop}{pcov}
\newcommand{\twopcov}[2]{\pcovop\bigl[#1,#2\bigr]}
\newcommand{\pcov}[1]{\pcovop\bigl[#1\bigr]}

\newcommand{\cf}{\varphi}
\newcommand{\scf}{\psi}
\newcommand{\df}{F}
\newcommand{\pdf}{f}
\newcommand{\cfunc}{h}
\DeclareMathOperator{\diagop}{diag}
\newcommand{\diag}[1]{\diagop(#1)}
\DeclareMathOperator{\rankop}{rank}
\newcommand{\rank}[1]{\rankop\bigl(#1\bigr)}
\newcommand{\dete}[1]{\det\bigl(#1\bigr)}
\DeclareMathOperator*{\diffop}{\Delta}
\newcommand{\difference}[3]{\diffop_{#1}^{#2}\bigl[#3\bigr]}

\newcommand{\sounumber}{m}
\newcommand{\altsounumber}{n}
\newcommand{\mixnumber}{p}

\newcommand{\matrixtype}[1]{\boldsymbol{#1}}

\DeclareMathOperator*{\specop}{\matrixtype{\lambda}}
\newcommand{\spectrum}[1]{\specop\bigl[#1\bigr]}
\newcommand{\speccomp}{\lambda}

\newcommand{\mixing}{\matrixtype{A}}
\newcommand{\mixingcol}{\boldsymbol{\alpha}}
\newcommand{\mixingel}{\alpha}
\newcommand{\altmixing}{\matrixtype{B}}
\newcommand{\altmixingcol}{\boldsymbol{\beta}}
\newcommand{\altmixingel}{\beta}
\newcommand{\compmatrix}{\matrixtype{C}}
\newcommand{\compmatrixtwo}{\matrixtype{D}}
\newcommand{\realm}{\compmatrix_R}
\newcommand{\imagm}{\compmatrix_I}
\newcommand{\solution}{\matrixtype{W}}
\newcommand{\permmatrix}{\matrixtype{P}}
\newcommand{\diagmatrix}{\matrixtype{\Lambda}}
\newcommand{\unitmatrix}{\matrixtype{U}}
\newcommand{\unittwomatrix}{\matrixtype{V}}
\newcommand{\orthmatrix}{\matrixtype{O}}
\newcommand{\identmatrix}{\matrixtype{I}}
\newcommand{\multizero}{\matrixtype{0}}
\newcommand{\mean}{\matrixtype{\mu}}

\DeclareMathOperator*{\Arg}{Arg}

\title{Complex Random Vectors and ICA Models: Identifiability, Uniqueness and Separability}

\author{Jan~Eriksson~\IEEEmembership{Member,~IEEE} and Visa~Koivunen~\IEEEmembership{Senior Member,~IEEE}
\thanks{Manuscript received March xx, 2004; revised December xx, 2005.
This work was supported in part by the Academy of Finland
and GETA Graduate School.}
\thanks{The authors are with the SMARAD CoE, Signal Processing Laboratory,
        Department of Electrical Engineering, 
        Helsinki University of Technology, FIN-02015 HUT, Finland
        (e-mail: \{jan.eriksson,visa.koivunen\}@hut.fi).}
}

\begin{document}
\maketitle

\begin{abstract}
In this paper the conditions for identifiability,
separability and uniqueness of linear complex
valued independent component analysis (ICA)  models are established.
These results extend the  well-known conditions for solving real-valued
ICA problems
to complex-valued models.
Relevant properties of complex random vectors are described in order to
extend the Darmois-Skitovich
theorem for complex-valued models. This theorem is used to construct a
proof
of a theorem for each of the above ICA model concepts.
Both circular and noncircular complex random vectors are covered.
Examples clarifying the above concepts are presented.
\end{abstract}


\section{Introduction}\label{sec:intro}

Independent component analysis (ICA)\cite{Comon:1994} is a relatively new
signal processing and data analysis technique. It may be used, for
example, in blind source separation (BSS)
and identifying or equalizing instantaneous multiple-input multiple-output
(I-MIMO) models.
It has found applications, e.g., in wireless communications,
biomedical signal processing and data mining 
(see \cite{Hyvarinen:2001} for references).
In instantaneous complex-valued ICA
problem 
\begin{equation}\label{eqn:complex_ica_intro}
  \mixture=\mixing\source,
\end{equation}
the goal is to recover the original source signal vectors $\source$
from the observation vectors $\mixture$ blindly without explicit knowledge of the sources or the linear 
mixing system $\mixing$.
ICA is based on the crucial assumption that the underlying unknown
source signals  are statistically independent.
Recent textbooks provide an interesting tutorial material and a partial review on ICA
 \cite{Cichocki:2002,Hyvarinen:2001}.

The theorems for linear combinations of real-valued random vectors and
theoretical conditions on separation
for real-valued signals are now well-known
\cite{Kagan:1973,Comon:1994,Eriksson:2004}.
Even though algorithms for separation of complex-valued
signals have been developed, for example \cite{Cardoso:1993,Comon:1994},
the conditions when the separation is possible have not been established.
Also recent papers, e.g.,
\cite{Bingham:2000,Calhoun:2002,Fiori:2003,Anemuller:2003},
proposing ICA algorithms for complex-valued data ignore this important
issue.

In this paper we construct theorems stating the conditions
for identifiability, separability, and uniqueness of complex-valued linear
ICA models.
These results extend the theorems proved for the real-valued instantaneous ICA
model
\cite{Comon:1994,Eriksson:2004} to the complex case.
Both circular (proper) and noncircular complex random vectors are covered
by the theorems.
These  conditions
depend not only on the probabilistic structure of the sources but also the
linear space structure of the mixing.
In order to prove the theorems, the celebrated Darmois-Skitovich theorem
\cite{Kagan:1973}
needs to be
extended to linear combinations of  complex random variables. A good number
of statistical properties of
circular and noncircular complex vectors have  to be  considered in the
process of constructing the proof.
This is due to the special operator structure that may be used for complex
random vectors. In addition, the
second order statistical properties of noncircular complex vectors
may not be defined using the covariance matrix alone
\cite{Neeser:1993,Picinbono:1996,Schreier:2003}.
General complex Gaussian random vectors is an important class of random
vectors that need to be addressed in detail.
There are relatively few papers
where noncircular complex random vectors are studied
\cite{Neeser:1993,Picinbono:1996,Vakhania:1996,Picinbono:1997,Grellier:2002,Schreier:2003}.
Hence, many of the key
results needed in proving the theorems are included in this paper and
presented in a unified manner. This also allows a direct derivation
of some fundamental information-theoretic quantities like the entropy
of a complex normal random vector.

The paper is organized as follows.
In Section~\ref{sec:crvcs} relevant properties
that distinguish complex random
vectors from real random vectors are described in detail.
Especially, the correlation structure
is used to study complex normal random vectors. These properties are
needed in proving the
Darmois-Skitovich theorem for the complex case. This theorem plays a key
role in establishing
the conditions for identifiablity, separability and uniqueness of complex
linear ICA models 
in Section~\ref{sec:complex_ica}.
Finally, some concluding remarks are given. Most of the proofs are
presented in 
appendices.

\section{Relevant Properties of Complex random vectors}\label{sec:crvcs}

The traditional probability theory is concerned with real-valued random variables (r.v.s) and
random vectors (r.vc.s). The theory has been generalized to various algebraic structures. 
Main studies are  in the frameworks of locally compact spaces and complete separable metric spaces
(see, e.g., \cite{Vakhania:1987,Ruzsa:1988,Dudley:1989,Feldman:1993} and references therein). 
However, the most
natural extension from the engineering point of view is the complex Hilbert space.
It seems  to have gained relatively little attention. Some results on
complex normal r.vc.s can be found in \cite{Wooding:1956,Goodman:1963}. The
second-order structure of complex r.vc.s has been studied in \cite{Neeser:1993,Picinbono:1996,Picinbono:1997,Schreier:2003}, and  a general
framework for higher-order statistics can be found from \cite{Amblard:1996}.
Some research has been conducted on complex elliptically symmetric
distributions \cite{Krishnaiah:1986}
and on complex stable distributions \cite{Hudson:2001}.
Polya's theorem to complex case is presented in \cite{Vakhania:1997}.
The only systematic Hilbert space approach known to the authors is \cite{Vakhania:1996}. 
This may be due to the fact that the additive structure of the complex Hilbert space
is the same as that of the real Hilbert space. However, the multiplicative structure and the operator
structure are different giving r.vc.s in a complex Hilbert space distinct 
properties. Even though many results from the general abstract theory apply
directly to the complex Hilbert space case, the systematic treatment considering both
the additive and the multiplicative structure seems to be missing.

In Section~\ref{ssc:isomorphism} the finite dimensional Hilbert space is reviewed
by constructing an isomorphism into a real-valued Hilbert space. This isomorphism shows
essentially the difference between the real and complex Hilbert spaces. %
In Section~\ref{ssc:crvc} some basic properties of r.vc.s in the complex Hilbert
space are stated, the second-order structure of complex r.vc.s is studied in 
Section~\ref{ssc:sorvc}. Complex normal r.vc.s are studied is Section~\ref{ssc:cGrvc}
and, finally the complex Darmois-Skitovich theorem is proved in Section~\ref{ssc:cDSt}.

\subsection{Notation}\label{ssc:notation}

Let us begin with some definitions and notations.
We have used typewriter font for all random objects, e.g. $\mixturecomp$, in order to distinguish
them from deterministic ones, e.g. $x$. For random vectors, e.g. $\mixture$, we have used the vec symbol 
in order to separate them from scalar random variables. For deterministic objects, the bold face
lower case letters are used for vectors, e.g. $\compvector$, and the bold face upper case letters are used for matrices,
e.g. $\solution.$

The \emph{modulus} of a complex number
$\cocomp=\realz+\jmath\imagz\in\mathbb{C}$ is denoted 
$|\cocomp|=\sqrt{\cconj{\cocomp}\cocomp}=\sqrt{\realz^2+\imagz^2}$, where the superscript $\cconj{}$ denotes
the complex \emph{conjugate}, $\cconj{\cocomp}=\realz-\jmath\imagz$, and $\jmath=\sqrt{-1}$ is the
imaginary unit. Recall that any nonzero complex number $\cocomp$ can be given in \emph{polar form}
$\cocomp=\rconst e^{\jmath\cargu}$, where $\rconst>0,\cargu\in\mathbb{R}$. The number $\cargu$
is called an \emph{argument} of the complex number $\cocomp$, and the argument $\cargu=\Arg(\cocomp)$ 
such that $-\pi\leq\cargu<\pi$ is called the \emph{principal argument}.
The real part of a $\mixnumber$-dimensional complex vector 
$(\cocomp_1\ \cocomp_2\ \cdots\  \cocomp_{\mixnumber})^\trans=\compvector\in\mathbb{C}^\mixnumber$,
where $\trans$ is the ordinary \emph{transpose},
is denoted by $\realvz$ and the imaginary part by $\imagvz$.
The Euclidean norm of a vector $\compvector$ is denoted 
$\norm{\compvector}^2=\inner{\compvector}{\compvector}=\compvector^\htrans\compvector$, where
$\inner{\cdot}{\cdot}$ is the inner  product and the superscript $\htrans$ denotes
the \emph{conjugate transpose}, i.e., the Hermitian adjoint. A complex matrix $\compmatrix\in\mathbb{C}^{\mixnumber\times\mixnumber}$
is termed \cite{Horn:1985} \emph{symmetric} if $\compmatrix^\trans=\compmatrix$
and \emph{Hermitian} if $\compmatrix^\htrans=\compmatrix$.  Furthermore, the matrix
$\compmatrix$ is \emph{orthogonal}
if $\compmatrix^\trans\compmatrix=\compmatrix\compmatrix^\trans=\identmatrix_{\mixnumber}$
and \emph{unitary}
if $\compmatrix^\htrans\compmatrix=\compmatrix\compmatrix^\htrans=\identmatrix_{\mixnumber}$, where
$\identmatrix_{\mixnumber}$ denotes the $\mixnumber\times\mixnumber$ identity matrix. 

\subsection{Complex Hilbert space isomorphism}\label{ssc:isomorphism}

Let  $\compmatrix=\realm+\jmath\imagm\in\mathbb{C}^{\sounumber\times\mixnumber}$ and
$\compvector=\realvz+\jmath\imagvz\in\mathbb{C}^\mixnumber$. We use 
the following notations 
\begin{equation} \
  \isoreal{\compmatrix}
  =\begin{pmatrix}
     \realm & -\imagm\\
     \imagm &  \realm
   \end{pmatrix}
  \text{ and } 
  \isoreal{\compvector}
   =\begin{pmatrix}
     \realvz \\
     \imagvz
   \end{pmatrix}
\end{equation}
for the associated $2\sounumber\times 2\mixnumber$ real matrix and $2\mixnumber$-variate real vector, 
respectively. The mapping $\compvector\mapsto\isoreal{\compvector}$ gives naturally a group isomorphism
between the additive Abelian groups $\mathbb{C}^\mixnumber$ and $\mathbb{R}^{2\mixnumber}$.
In the case $\sounumber=\mixnumber=1$, the mapping given by $\compmatrix\mapsto\isoreal{\compmatrix}$
defines a \emph{field isomorphism} (e.g., \cite{Goodman:1963,Vakhania:1996}) 
between the complex numbers and a subset of real two dimensional matrices. Therefore, one
can construct real structures where the role of complex multiplication is played
by the special matrices.

Now consider the mapping
\begin{equation}\label{eqn:isomorphism}
   \compmatrix\compvector\mapsto
     \isoreal{(\compmatrix\compvector)}=\isoreal{\compmatrix}\isoreal{\compvector}.
\end{equation}
It is continuous and therefore preserves the topological properties, i.e.,
it is a homeomorphism \cite{Dudley:1989}. 
Let $\diag{\compvector}$ (as in Matlab)
denote the \emph{diagonal matrix} with components of $\compvector$ in its main diagonal
and zeros elsewhere.
Since 
$\mathbb{C}^{\mixnumber}$ is a vector space, where the scalar multiplication 
for $\cconst\in\mathbb{C}$ is given by
\begin{equation}
 \cconst\compvector\triangleq
      \begin{pmatrix}
        \cconst\cocomp_1\\
        \vdots\\
        \cconst\cocomp_\mixnumber
      \end{pmatrix}
      =
      \diag{\begin{pmatrix} \cconst & \cdots & \cconst \end{pmatrix}}
      \compvector,
\end{equation}
the mapping (\ref{eqn:isomorphism}) defines a vector space isomorphism between the
standard $\mixnumber$-dimensional complex vector space and a $2\mixnumber$-dimensional real-valued vector space given by the mapping.
It is important to realize that this associated real-valued vector space 
is \emph{not} isomorphic to the standard real vector space 
$\mathbb{R}^{2\mixnumber}$. 
Furthermore, by equating $\compvector_1^\htrans$ with $\compmatrix$
in (\ref{eqn:isomorphism}) it is easily verified that the
mapping $\mathbb{C}\to\mathbb{R}^{2}:\ \compvector_1^\htrans\compvector_2\mapsto\isoreal{(\compvector_1^\htrans)}\isoreal{(\compvector_2)}$
associates a (complex) inner product for $\mathbb{R}^{2\mixnumber}$.
Therefore,
the mapping (\ref{eqn:isomorphism}) is also a Hilbert space isomorphism. Again, it should
be emphasized that the inner product given by the mapping is not the standard  Euclidean inner
product in $\mathbb{R}^{2\mixnumber}$. However, the vector norms, and hence metrics, are equivalent
in both.

The following
properties are easily established.
\begin{lemma}\label{lem:matrix_properties}
 Let $\compmatrix\in\mathbb{C}^{\mixnumber\times\mixnumber}$ and 
     $\compvector\in\mathbb{C}^\mixnumber$.
  \begin{enumerate}[(i)]
    \item $|\dete{\compmatrix}|^2=\dete{\isoreal{\compmatrix}}$.
    \item $\compmatrix$ is Hermitian iff $\isoreal{\compmatrix}$ is symmetric. Then 
          $\dete{\compmatrix}^2=\dete{\isoreal{\compmatrix}}$ and
          $2\times\rank{\compmatrix}=\rank{\isoreal{\compmatrix}}$.
    \item $\compmatrix$ is nonsingular iff $\isoreal{\compmatrix}$ is nonsingular.
    \item $\compmatrix$ is unitary iff $\isoreal{\compmatrix}$ is orthogonal.
    \item $\compvector^\htrans\compmatrix\compvector=
           \isoreal{\compvector}^\trans\isoreal{\compmatrix}\isoreal{\compvector}$ 
    \item $\compmatrix$ is Hermitian positive definite iff $\isoreal{\compmatrix}$ 
          is symmetric positive definitive.
    \item Any polynomial with complex coefficients in variables $\isoreal{\compvector}$
          can be equivalently given in variables $(\compvector,\cconj{\compvector})$. \label{eqn:poly_wide}
  \end{enumerate}
\end{lemma}
\begin{proof}
  These properties are direct consequences of the isomorphism, 
  see, e.g., \cite{Goodman:1963,Krishnaiah:1986}. The last property
  follows from the identities $\realvz=\frac12(\compvector+\cconj{\compvector})$ and
  $\imagvz=\frac{-\jmath}{2}(\compvector-\cconj{\compvector})$. 
\end{proof}

Since the variables $(\compvector,\cconj{\compvector})$ 
in Lemma~\ref{lem:matrix_properties}(\ref{eqn:poly_wide}) are dependent,
we call such complex polynomials 
 \emph{wide sense polynomials}. 
The idea of using also the complex conjugate
variable has turned out to be highly useful in, e.g., complex parameter estimation \cite{Picinbono:1995}
and blind channel equalization \cite{Grellier:2002}.

\subsection{Complex random vectors}\label{ssc:crvc}

A $\mixnumber$-variate complex random vector (r.vc.) $\mixture$ 
is defined as an r.vc. of the form
\begin{equation}
  \mixture=\realrvc+\jmath\imagrvc,
\end{equation}
where $\realrvc$ and $\imagrvc$ are $\mixnumber$-variate real r.vc.s, i.e., $\realrvc$
and $\imagrvc$ are measurable functions from a probability space to $\mathbb{R}^{\mixnumber}$.
This is equivalent
for $\mixture$ to be measurable from the probability space into $\mathbb{C}^{\mixnumber}$
due to the separability of the complex space.
Therefore, the probabilistic structure of the r.vc.s in $\mathbb{C}^{\mixnumber}$ and
the probabilistic structure of the r.vc.s in $\mathbb{R}^{2\mixnumber}$ is the same.
However, the \emph{operator structure is different} as it is evident from the
previous section. This gives distinct properties to the r.vc.s with complex
values, and justifies studying them separately. Throughout this paper all complex r.vc.s
are assumed to be \emph{full}. This means that the support of the induced measure of 
a $\mixnumber$-dimensional r.vc. is not contained in any lower dimensional complex subspace.

Since the probabilistic structures of r.vc.s in $\mathbb{C}^{\mixnumber}$ and in
$\mathbb{R}^{2\mixnumber}$ are the same, also the operator structure of r.vc.s
in  $\mathbb{C}^{\mixnumber}$ can be studied by first using the isomorphism
(\ref{eqn:isomorphism}) and then applying the concepts associated with the real
r.vc.s. However, we define these associated concepts directly on $\mathbb{C}^{\mixnumber}$,
since this approach is notationally more convenient. 

The \emph{expectation} $\Eop[\cdot]$ of a complex r.vc. $\mixture$ is defined as 
\begin{equation}
  \E{\mixture}{\mixture}=\E{\realrvc}{\realrvc}+\jmath\E{\imagrvc}{\imagrvc},
\end{equation}
and the \emph{distribution function} $\df_{\mixture}$ is given
as $\df_{\mixture}(\compvector)\triangleq\df_{\isoreal{\mixture}}(\isoreal{\compvector})$,
where 
$\compvector=(\cocomp_1,\ldots,\cocomp_\mixnumber)^\trans\in\mathbb{C}^\mixnumber$
and $\df_{\isoreal{\mixture}}$ denotes the distribution function of real-valued
r.vc. $\isoreal{\mixture}$.
Then for independent r.v.s 
$(\sourcecomp_1,\ldots,\sourcecomp_{\mixnumber})^\trans=\source$, we have
\begin{equation}\label{eqn:df_independence}
  \df_{\source}(\compvector)  =\df_{\isoreal{\source}}(\isoreal{\compvector})
     = \prod_{\dindex=1}^{\mixnumber}
       \df_{\isoreal{(\sourcecomp_\dindex)}}(\isoreal{(\cocomp_\dindex)})
     = \prod_{\dindex=1}^{\mixnumber}
       \df_{\sourcecomp_\dindex}(\cocomp_\dindex).
\end{equation}
The same way we define the probability \emph{density function} $\pdf_{\mixture}$ (if it exists) 
of  a $\mixnumber$-dimensional complex r.vc. $\mixture$ as 
$\pdf_{\mixture}(\compvector)\triangleq\pdf_{\isoreal{\mixture}}(\isoreal{\compvector})$,
and the \emph{characteristic function} (c.f.) \cite{Vakhania:1996} as
\begin{equation}
\begin{split}
  \cf_{\mixture}(\compvector) 
    \triangleq \cf_{\isoreal{\mixture}}(\isoreal{\compvector})
   = &\E{\isoreal{\mixture}}{\exp\bigl(\jmath\inner{\isoreal{\compvector}}{\isoreal{\mixture}}\bigr)}\\
   = &\E{\mixture}{\exp\bigl(\jmath\real{\inner{\compvector}{\mixture}}\bigr)}.
 \end{split}
\end{equation}
It follows directly from Eq.~(\ref{eqn:df_independence})
that for independent complex r.v.s $(\sourcecomp_1,\ldots,\sourcecomp_\mixnumber)^\trans=\source$,
\begin{equation}\label{eqn:indi}  
  \cf_{\source}(\compvector)
         =\prod_{k=1}^\mixnumber\cf_{\sourcecomp_\dindex}(\cocomp_\dindex).
\end{equation}
Using a standard property of real c.f.s 
and the properties of the isomorphism~(\ref{eqn:isomorphism}), 
we have a useful relation for the c.f. of an r.vc. $\mixture$ and the c.f. of the linearly transformed
r.vc. $\compmatrix\mixture$. Namely, 
for any complex matrix $\compmatrix$, we have
\begin{equation}\label{eqn:htrans}
\begin{split}
  \cf_{\compmatrix\mixture}(\compvector)
  = &\cf_{\isoreal{(\compmatrix\mixture)}}(\isoreal{\compvector})
  =\cf_{\isoreal{\compmatrix}\isoreal{\mixture}}(\isoreal{\compvector})
  =\cf_{\isoreal{\mixture}}((\isoreal{\compmatrix})^\trans\isoreal{\compvector})\\
  = &\cf_{\isoreal{\mixture}}(\isoreal{(\compmatrix^\htrans)}\isoreal{\compvector})
  =\cf_{\isoreal{\mixture}}(\isoreal{(\compmatrix^\htrans\compvector)})
  =\cf_{\mixture}(\compmatrix^\htrans\compvector).
\end{split}
\end{equation}
Finally, a c.f. $\cf_{\mixture}(\compvector)$ is called \emph{analytic} if
$\cf_{\isoreal{\mixture}}(\isoreal{\compvector})$ is an analytic c.f. \cite{Cuppens:1975},
i.e., the real c.f. $\cf_{\isoreal{\mixture}}(\isoreal{\compvector})$ has a regular extension
defined on $\mathbb{C}^{2\mixnumber}$ in some neighborhood of the origin.

\subsection{Second-order statistics of complex random vectors}\label{ssc:sorvc}

An r.vc. $\mixture$ has \emph{finite second order} or \emph{weak second order} \cite{Vakhania:1996} statistics
if $\E{\mixture}{|\inner{\mixture}{\compvector}|^2}<\infty$ for all 
$\compvector\in\mathbb{C}^\mixnumber$. This is clearly equivalent to the
existence of finite second order statistics for both real r.vc.s $\realrvc$ and $\imagrvc$.
All r.vc.s in this section are assumed to have finite second order statistics. Such r.vc.s
are in general called \emph{second-order} complex r.vc.s.

The second-order statistics between two real r.vc.s may be described by the covariance matrix.
The complex \emph{covariance matrix} 
$\twocov{\mixture_1}{\mixture_2}$ of two complex r.vc.s $\mixture_1$ and $\mixture_2$ may be
defined as
\begin{equation}\label{def:cov}
  \twocov{\mixture_1}{\mixture_2}\triangleq
    \E{\mixture_1,\mixture_2}{(\mixture_1-\E{\mixture_1}{\mixture_1})
                              (\mixture_2-\E{\mixture_2}{\mixture_2})^\htrans}.
\end{equation}
However, considering the real representations of the complex r.vc.s, it can be seen that
the complex covariance matrix does not give complete second order description. For that we define
the \emph{pseudo-covariance matrix}\footnote{The pseudo-covariance matrix is called 
the \emph{relation matrix} in \cite{Picinbono:1996} 
and the \emph{complementary covariance matrix} in \cite{Schreier:2003}.} 
$\twopcov{\mixture_1}{\mixture_2}$ \cite{Neeser:1993} as
\begin{equation}\label{def:pcov}
\begin{split}
  \twopcov{\mixture_1}{\mixture_2} \triangleq &
    \E{\mixture_1,\mixture_2}{(\mixture_1-\E{\mixture_1}{\mixture_1})
                              (\mixture_2-\E{\mixture_2}{\mixture_2})^\trans}\\
                             = &\twocov{\mixture_1}{\cconj{\mixture_2}}.
\end{split}
\end{equation}
Two complex r.vc.s $\mixture_1$ and $\mixture_2$ are
\emph{uncorrelated} if real r.vc.s 
$\isoreal{(\mixture_1)}$ and $\isoreal{(\mixture_2)}$ are uncorrelated, i.e.,
$\twocov{\isoreal{(\mixture_1)}}{\isoreal{(\mixture_2)}}=\multizero_{2\mixnumber\times2\mixnumber}$, where 
$\multizero_{2\mixnumber\times2\mixnumber}$
denotes the $2\mixnumber\times2\mixnumber$ matrix of zeros. Then, by using the properties from the previous section,
the following lemma \cite{Neeser:1993} follows directly.
\begin{lemma}\label{thm:twounc}
  Complex r.vc.s $\mixture_1$ and $\mixture_2$ are uncorrelated if and only if
  $\twocov{\mixture_1}{\mixture_2}=\twopcov{\mixture_1}{\mixture_2}=\multizero_{\mixnumber\times\mixnumber}$.
\end{lemma}

As it is the case with real r.vc.s, 
the internal correlation structure of a single r.vc. $\mixture$ may be of interest
in addition to correlation between two r.vc.s.
Then we define $\cov{\mixture}\triangleq\twocov{\mixture}{\mixture}$ and
$\pcov{\mixture}\triangleq\twopcov{\mixture}{\mixture}$, and call them
the covariance matrix and the pseudo-covariance matrix of an r.vc. $\mixture$,
respectively.
It is easily seen that the covariance matrix $\cov{\mixture}$ is Hermitian
and the pseudo-covariance matrix is symmetric. Since all r.vc.s are assumed to be
full, the covariance matrix $\cov{\mixture}$ is also positive definite.
R.vc. $\mixture$ is said to have \emph{uncorrelated components} if 
all its marginal r.v.s $\mixturecomp_\dindex$ and $\mixturecomp_\dindextwo$, 
$\dindex\neq\dindextwo$,
are uncorrelated. The following lemma is a simple consequence of Lemma~\ref{thm:twounc}.
\begin{lemma}\label{thm:unccom}
  A complex r.vc. $\mixture$ has uncorrelated components if and only if
  its covariance matrix and pseudo-covariance matrix are diagonal.
\end{lemma}

An r.vc. $\mixture$ is said to be spatially \emph{white}, if 
$\cov{\mixture}=\sigma^2\identmatrix_{\mixnumber}$ for some $\sigma^2>0$. 
If
$\pcov{\mixture}=\multizero_{\mixnumber\times\mixnumber}$, then the r.vc. is called 
\emph{second order circular} (or circularly symmetric).
Some authors prefer the term \emph{proper} \cite{Neeser:1993,Vakhania:1996}.
Circular r.vc.s have gained most of the attention in the literature of complex r.vc.s. This is likely
due to the fact that all the second order information of circular r.vc.s is contained
in the covariance matrix, which, on the other hand, behaves like the covariance
matrix for the real r.vc.s. However, in this paper
we need the complete second-order description to
be derived next. Our approach is to our best knowledge novel, mainly based on the
following theorem. For alternative characterizations, see 
\cite{Picinbono:1996,Vakhania:1996,Schreier:2003}.

\begin{lause}\label{thm:compdef}
  Any full complex $\mixnumber$-dimensional r.vc. $\mixture$ with finite second order statistics
  can be transformed by using a nonsingular square matrix $\compmatrix$ such that the r.vc.
  $\source=(\sourcecomp_1,\ldots,\sourcecomp_{\mixnumber})^\trans
  =\compmatrix\mixture$ has the following properties:
  \begin{enumerate}[(i)]
    \item $\cov{\source}=\identmatrix_{\mixnumber}$
    \item $\pcov{\source}=\diag{\spectrum{\source}}$, where
          $\spectrum{\source}=(\speccomp_1,\ldots,\speccomp_{\mixnumber})^\trans$ denotes
          a vector such that 
          $\speccomp_1\geq\cdots\geq\speccomp_{\mixnumber}$.         
  \end{enumerate}
\end{lause}
\begin{proof}
  It is easily verified that $\cov{\compmatrix\mixture}=\compmatrix\cov{\mixture}\compmatrix^\htrans$
  and $\pcov{\compmatrix\mixture}=\compmatrix\pcov{\mixture}\compmatrix^\trans$. 
  By Corollary $4.6.12(b)$ in \cite{Horn:1985}, if a matrix $\mixing$ is Hermitian and positive definite
  and a matrix $\altmixing$ is symmetric, then there exists a nonsingular 
  matrix $\compmatrix$ such that $\compmatrix\mixing\compmatrix^\htrans=\identmatrix_{\mixnumber}$ and
  $\compmatrix\altmixing\compmatrix^\trans$ is a diagonal matrix with nonnegative diagonal entries.
  Since the covariance matrix is Hermitian and positive definitive and the pseudo-covariance
  matrix is symmetric, the proof is completed by noticing that 
  the diagonal entries can be ordered by permutating the rows of $\compmatrix$.
\end{proof}

Since  $\cov{\mixturecomp}=\cov{\realpart{\mixturecomp}}+\cov{\imagpart{\mixturecomp}}$    
   and     $\pcov{\mixturecomp}
         =\cov{\realpart{\mixturecomp}}-\cov{\imagpart{\mixturecomp}}
          +2\jmath\twocov{\realpart{\mixturecomp}}{\imagpart{\mixturecomp}}$ for any complex
  r.v. $\mixturecomp=\realpart{\mixturecomp}+\jmath\imagpart{\mixturecomp}$,  it follows
that in Theorem~\ref{thm:compdef}
 $\twocov{\real{\sourcecomp_{\dindex}}}{\imag{\sourcecomp_{\dindex}}}=0$ and
          $1\geq\speccomp_{\dindex}=\cov{\real{\sourcecomp_{\dindex}}}-
           \cov{\imag{\sourcecomp_{\dindex}}}\geq0$, $\dindex=1,\ldots,\mixnumber$.
The r.vc.s satisfying the properties of Theorem~\ref{thm:compdef} have a special structure,
and they are here called \emph{strongly uncorrelated}. 
Any strongly uncorrelated r.vc. is white with $\cov{\source}=\identmatrix_{\mixnumber}$,
but the converse is not true. 
In general, for a given r.vc. $\mixture$, the strongly uncorrelated r.vc. $\source$
and the \emph{strong-uncorrelating} transform $\compmatrix$ given by Theorem~\ref{thm:compdef}
are not unique. However, we have the following.

\begin{lause}\label{thm:compuni}
   For a given r.vc. $\mixture$, the vector $\spectrum{\source}$ in Theorem~\ref{thm:compdef}
   is unique. 
\end{lause}
\begin{proof}
 Suppose there exist two nonsingular transformations $\compmatrix_1$ and $\compmatrix_2$ such
 that r.vc.s $\source_1=\compmatrix_1\mixture$ and $\source_2=\compmatrix_2\mixture$ satisfy
 the properties in Theorem~\ref{thm:compdef}. Let 
 $\compmatrix_1=\unitmatrix_1\diagmatrix_1\unittwomatrix_1^\htrans$ and
 $\compmatrix_2=\unitmatrix_2\diagmatrix_2\unittwomatrix_2^\htrans$ be the
 singular value decompositions (SVD) (see \cite{Horn:1985}) of the transform matrices.
 Now 
 $\identmatrix_{\mixnumber}=\compmatrix_1\cov{\mixture}\compmatrix_1^\htrans
              =\compmatrix_2\cov{\mixture}\compmatrix_2^\htrans$, and therefore
 $\cov{\mixture}=\unittwomatrix_1\diagmatrix_1^{-2}\unittwomatrix_1^\htrans
                =\unittwomatrix_2\diagmatrix_2^{-2}\unittwomatrix_2^\htrans$.
 Since $\cov{\mixture}$ is positive definite, it follows 
 $\unittwomatrix_1\diagmatrix_1\unittwomatrix_1^\htrans
 =\unittwomatrix_2\diagmatrix_2\unittwomatrix_2^\htrans$.
 Now 
 \begin{equation}\label{eqn:pcov_TF}
 \begin{split}
  \pcov{\source_1} = &\unitmatrix_1\diagmatrix_1\unittwomatrix_1^\htrans\pcov{\mixture}
                      \cconj{\unittwomatrix_1}\diagmatrix_1\unitmatrix_1^\trans\\
                   = &\unitmatrix_1(\unittwomatrix_1^\htrans\unittwomatrix_1)
                           \diagmatrix_1\unittwomatrix_1^\htrans\pcov{\mixture}
                      \cconj{\unittwomatrix_1}\diagmatrix_1
                      (\unittwomatrix_1^\trans\cconj{\unittwomatrix_1})
                      \unitmatrix_1^\trans\\
                   = &\unitmatrix_1\unittwomatrix_1^\htrans(\unittwomatrix_1
                           \diagmatrix_1\unittwomatrix_1^\htrans)\pcov{\mixture}
                      (\cconj{\unittwomatrix_1}\diagmatrix_1
                      \unittwomatrix_1^\trans)\cconj{\unittwomatrix_1}
                      \unitmatrix_1^\trans\\
                   = &\unitmatrix_1\unittwomatrix_1^\htrans(\unittwomatrix_2
                           \diagmatrix_2\unittwomatrix_2^\htrans)\pcov{\mixture}
                      (\cconj{\unittwomatrix_2}\diagmatrix_2
                      \unittwomatrix_2^\trans)\cconj{\unittwomatrix_1}
                      \unitmatrix_1^\trans\\
                   = &\unitmatrix_1\unittwomatrix_1^\htrans\unittwomatrix_2
                           (\unitmatrix_2^\htrans\unitmatrix_2)
                           \diagmatrix_2\unittwomatrix_2^\htrans\pcov{\mixture}\\
                   & \cconj{\unittwomatrix_2}\diagmatrix_2
                           (\unitmatrix_2^\trans\cconj{\unitmatrix_2})
                      \unittwomatrix_2^\trans\cconj{\unittwomatrix_1}
                      \unitmatrix_1^\trans\\
                   = &\unitmatrix_1\unittwomatrix_1^\htrans\unittwomatrix_2
                           \unitmatrix_2^\htrans(\unitmatrix_2
                           \diagmatrix_2\unittwomatrix_2^\htrans\pcov{\mixture}\\
                     & \cconj{\unittwomatrix_2}\diagmatrix_2
                            \unitmatrix_2^\trans)\cconj{\unitmatrix_2}
                      \unittwomatrix_2^\trans\cconj{\unittwomatrix_1}
                      \unitmatrix_1^\trans\\
                   = &\unitmatrix_1\unittwomatrix_1^\htrans\unittwomatrix_2
                           \unitmatrix_2^\htrans
                           \pcov{\source_2}\cconj{\unitmatrix_2}
                      \unittwomatrix_2^\trans\cconj{\unittwomatrix_1}
                      \unitmatrix_1^\trans,                
 \end{split}
 \end{equation}
 and since $\unitmatrix_1\unittwomatrix_1^\htrans\unittwomatrix_2\unitmatrix_2^\htrans$ 
 is unitary, $\pcov{\source_1}$ and $\pcov{\source_2}$ have the same singular values.
 Since by the assumption  $\pcov{\source_1}$ and $\pcov{\source_2}$ are diagonal
 with sorted entries,
 it follows $\pcov{\source_1}=\pcov{\source_2}$.
\end{proof}

\begin{huom}
  The proof of Theorem~\ref{thm:compuni} gives a way to construct a strong-uncorrelating
  transform $\compmatrix$ as follows:
  \begin{enumerate}[(i)]
    \item Find the usual whitening transform $\compmatrixtwo=\cov{\mixture}^{-\frac12}$, 
          i.e., the inverse of the matrix square root of $\cov{\mixture}$.
    \item Any symmetric matrix $\altmixing$ has a special form
          of SVD known as \emph{Takagi's factorization} (see \cite{Horn:1985}). 
          The factorization is given
          as $\altmixing=\unitmatrix\diagmatrix\unitmatrix^\trans$,
          where $\unitmatrix$ is unitary and $\diagmatrix$ is a diagonal
          matrix with real nondecreasing nonnegative main diagonal entries. 
          An example of the factorization is given in Eq.~(\ref{eqn:pcov_TF}).
          Hence, find
          $\pcov{\compmatrixtwo\mixture}=\unitmatrix\diagmatrix\unitmatrix^\trans$.
    \item Set $\compmatrix=\unitmatrix^\htrans\compmatrixtwo$.
  \end{enumerate}
  Notice also that the vector $\spectrum{\source}$ contains the singular values of the
  pseudo-covariance matrix of a white r.vc. with unit variances.
\end{huom}

The previous theorems lead to a useful characterization of second-order complex r.vc.s.

\begin{maar}
  The vector
  $\spectrum{\mixture}\triangleq\spectrum{\source}=(\speccomp_1,\ldots,\speccomp_{\mixnumber})^\trans$ 
  in Theorem~\ref{thm:compdef} is called the \emph{circularity spectrum} of an r.vc. $\mixture$.
  An element of the circularity spectrum corresponding to an r.v. is called a \emph{circularity coefficient}. 
\end{maar}

Any r.vc. $\mixture$ is clearly second order circular if and only if its circularity spectrum is a zero vector,
i.e., $\spectrum{\mixture}=\multizero_{\mixnumber\times1}$.

\begin{seur}\label{thm:unique_swhite}
  If the circularity spectrum of an r.vc. has distinct elements, all rows corresponding to nonzero
  circularity coefficients of the
  strong-uncorrelating transform are unique up to multiplication of the row by $-1$.
  A row corresponding to the zero
  coefficient is unique up to 
  multiplication of the row by $e^{\jmath\cargu}$, $\cargu\in\mathbb{R}$.
\end{seur}
\begin{proof}
 The left unitary factor in the SVD of a block matrix with distinct singular values is determined
 up to right multiplication by the matrix 
 $\diagmatrix=\diag{e^{\jmath\cargu_1},\ldots,e^{\jmath\cargu_\mixnumber}}$
 and the right unitary factor is determined by the left unitary factor \cite{Horn:1985}.
 In the special form for a symmetric matrix (Takagi's factorization), 
 $\cargu_\dindex=0$ or $\cargu_\dindex=\pi$ for
 the values of $\dindex$ corresponding to nonzero singular values. Therefore, 
 $\unitmatrix_1\unittwomatrix_1^\htrans\unittwomatrix_2\unitmatrix_2^\htrans=\diagmatrix$
 in Eq.~(\ref{eqn:pcov_TF}), and 
 \begin{equation}
 \begin{split}
   \compmatrix_1= &\unitmatrix_1\diagmatrix_1\unittwomatrix_1^\htrans
                  =\unitmatrix_1\unittwomatrix_1^\htrans
                  (\unittwomatrix_1\diagmatrix_1\unittwomatrix_1^\htrans)\\
                = &\diagmatrix\unitmatrix_2\unittwomatrix_2^\htrans
                 (\unittwomatrix_2\diagmatrix_2\unittwomatrix_2^\htrans)
                =\diagmatrix\unitmatrix_2\diagmatrix_2\unittwomatrix_2^\htrans
                =\diagmatrix\compmatrix_2
 \end{split}
 \end{equation}
 by the proof of Theorem~\ref{thm:compuni}.
\end{proof}

Some properties of the circularity coefficient are listed in the following lemma, whose proof is given
in Appendix~\ref{sec:proofcclemma}.

\begin{lemma}\label{thm:cc_properties}
  Let $\mixturecomp$ and $\solucomp$ be uncorrelated  second-order complex r.v.s. Then
  \begin{enumerate}[(i)]
    \item $0\leq\spectrum{\cconst\mixturecomp}=\spectrum{\mixturecomp}
          =\frac{|\pcov{\mixturecomp}|}{\cov{\mixturecomp}}\leq1$ for any nonzero 
          constant $\cconst\in\mathbb{C}$,
    \item $\spectrum{\mixturecomp}=1$ if and only if 
          $\mixturecomp=\cconst(\realpart{\sourcecomp}+\jmath\rconst)$ for some unit variance real
          r.v. $\realpart{\sourcecomp}$ and deterministic constants $0\neq\cconst\in\mathbb{C}$,
          $\rconst\in\mathbb{R}$,
    \item $\spectrum{\mixturecomp+\solucomp}=
           \frac{|\pcov{\mixturecomp}+\pcov{\solucomp}|}{\cov{\mixturecomp}+\cov{\solucomp}}\leq
           \max\{\spectrum{\mixturecomp},\spectrum{\solucomp}\}$
           with the equality if and only if $\spectrum{\mixturecomp}=\spectrum{\solucomp}$ and
           $\Arg(\pcov{\mixturecomp})=\Arg(\pcov{\solucomp})$ if  $\spectrum{\mixturecomp}\neq0$.
  \end{enumerate}
\end{lemma}

\subsection{Complex normal random vectors}\label{ssc:cGrvc}

There are no commonly agreed definitions of what is meant by complex normal r.vc.s.
It is natural to require that a r.vc. $\mixture$ is normal (Gaussian) if
the real r.vc. $\isoreal{\mixture}$ is multivariate normal. 
Such r.vc.s are
generally called \emph{wide sense normal} r.vc.s \cite{Vakhania:1996}. 
Since the real complex normal r.vc. is completely characterized by its
mean vector and covariance, the results from the previous section
show that a wide sense complex normal r.vc. is completely specified
by its mean, covariance matrix, and pseudo-covariance matrix.

However,
all wide sense normal r.vc.s do not possess all the properties that real normal
r.vc.s do. Only a special subclass of wide sense normal r.vc.s has a
density function similar to the real r.vc.s \cite{Wooding:1956,Goodman:1963},
maximizes the entropy \cite{Neeser:1993}, or has the 2-stability property 
(Polya's characterization) \cite{Vakhania:1997}.
Such r.vc.s are called \emph{narrow sense normal} r.vc.s \cite{Vakhania:1996}. 
They are wide sense normal r.vc.s such that the real and imaginary
parts of any linear projection of the r.vc. are independent and have equal
variances. This condition is equivalent
to the requirement that a wide sense normal r.vc. is second order circular (see, e.g., \cite{Neeser:1993}).

In order to establish the properties of the complex 
ICA model of Eq.~(\ref{eqn:complex_ica_intro}), neither
wide sense normal in its full generality nor narrow sense normal is
adequate, and a more specific characterization of complex normal r.vc.s
is needed. This is done next.
From now on, we will use the term ``complex normal'' to mean wide sense complex
normal r.vc.

The main result is the following decomposition theorem for complex normal random vectors.

\begin{lause}\label{thm:normal_rep}
  An r.vc. $\mnormal$ is complex normal with circularity spectrum $\specop$ if and only if 
  \begin{equation}\label{eqn:stand_CG}
    \mnormal=\compmatrix(\realpart{\smnormal}+\jmath\imagpart{\smnormal})+\mean
  \end{equation}
  for some nonsingular matrix $\compmatrix$, a complex constant vector $\mean$, and 
  multinormal real independent r.vc.s 
  $\realpart{\smnormal}\sim 
  N\bigl(\multizero_{\mixnumber\times1},\frac12\identmatrix_{\mixnumber}+\frac12\diag{\specop}\bigr)$ 
  and $\imagpart{\smnormal}\sim 
  N\bigl(\multizero_{\mixnumber\times1},\frac12\identmatrix_{\mixnumber}-\frac12\diag{\specop}\bigr)$.
  Also $\cov{\mnormal}=\compmatrix\compmatrix^\htrans$,
       $\pcov{\mnormal}=\compmatrix\diag{\specop}\compmatrix^\trans$, and
       $\E{\mnormal}{\mnormal}=\mean$.
\end{lause}
\begin{proof}
  It is obvious that the r.vc. $\mnormal$ in Eq.~(\ref{eqn:stand_CG}) is complex normal,
  $\cov{\mnormal}=\compmatrix\compmatrix^\htrans$, 
  $\pcov{\mnormal}=\compmatrix\diag{\specop}\compmatrix^\trans$, and
  $\E{\mnormal}{\mnormal}=\mean$. Thus, it remains to show that any complex normal r.vc.
  can be given the form (\ref{eqn:stand_CG}).
  
  Let $\mnormal$ be a complex normal r.vc. Without loss of generality assume it is zero mean. 
  By Theorem~\ref{thm:compdef}, there exists
  a nonsingular matrix $\compmatrixtwo$ such that $\cov{\compmatrixtwo\mnormal}=\identmatrix_{\mixnumber}$
  and $\pcov{\compmatrixtwo\mnormal}=\diag{\specop}$. Let 
  $\realpart{\smnormal}\sim N\bigl(\multizero_{\mixnumber\times1},\frac12\identmatrix_{\mixnumber}+\frac12\diag{\specop}\bigr)$ 
  and $\imagpart{\smnormal}\sim 
  N\bigl(\multizero_{\mixnumber\times1},\frac12\identmatrix_{\mixnumber}-\frac12\diag{\specop}\bigr)$ be real independent r.vc.s.
  Now 
  $\cov{\realpart{\smnormal}+\jmath\imagpart{\smnormal}}=
   \frac12\identmatrix_{\mixnumber}+\frac12\diag{\specop}+\frac12\identmatrix_{\mixnumber}-\frac12\diag{\specop}=\identmatrix_{\mixnumber}$ and
  $\pcov{\realpart{\smnormal}+\jmath\imagpart{\smnormal}}=\diag{\specop}$. 
  Hence $\compmatrixtwo\mnormal$ and $\realpart{\smnormal}+\jmath\imagpart{\smnormal}$ have the same
  second order structure.
   Since
   a zero mean complex normal r.vc. is completely characterized by the covariance
   and the pseudo-covariance matrices, it follows 
   $\compmatrixtwo\mnormal=\realpart{\smnormal}+\jmath\imagpart{\smnormal}$,
   and the claim follows by setting $\compmatrix=\compmatrixtwo^{-1}$.
\end{proof}

A complex normal r.vc. $\smnormal$ such that $\compmatrix=\identmatrix_{\mixnumber}$
and $\mean=\multizero_{\mixnumber\times1}$ in the representation (\ref{eqn:stand_CG}), i.e.,
$\smnormal=\realpart{\smnormal}+\jmath\imagpart{\smnormal}$, is
called \emph{standard complex normal} with the circularity spectrum $\specop$. 
Clearly any centered and strongly uncorrelated
complex normal r.vc. is standard.  Also, it is seen that any complex normal r.vc.
may be alternatively specified by the mean, the circularity spectrum, and the (inverse of) 
strong-uncorrelating matrix $\compmatrix$.

The previous decomposition allows the derivation of \emph{differential entropy} of 
a complex normal r.vc. in a closed form. Entropy $h(\mnormal)$
of an r.vc. $\mixture$ is defined as the entropy \cite{Cover:1991} of the 
real r.vc. $\isoreal{\mixture}$.
The following result has been implicitly derived in \cite{Taubock:2002} without
reference to circularity coefficients.

\begin{seur}\label{thm:entropy_normal}
  The differential entropy  $h(\mnormal)$ of a zero-mean complex normal 
  r.vc. $\mnormal$ with the circularity coefficients
   $\speccomp_{\dindex}\neq1$, $\dindex=1,\ldots,\mixnumber$, is given by
  \begin{equation}\label{eqn:entropy_normal}
    h(\mnormal)=\log\bigl(\det(\pi e\cov{\mnormal})\bigr)+
                \frac12\sum_{\dindex=1}^{\mixnumber}\log(1-\speccomp_{\dindex}^2).
  \end{equation}
\end{seur}
\begin{proof}
 Let $\mnormal=\compmatrix\smnormal$ be the decomposition 
 given by Theorem~\ref{thm:normal_rep}. 
 Now $\det(2\cov{\isoreal{\smnormal}})
 =\prod_{\dindex=1}^{\mixnumber}(1-\speccomp_{\dindex}^2)$,
 and the differential entropy of real-valued normal 
 r.vc. \cite{Cover:1991} simplifies as
 \begin{equation}
 \begin{split}
   h(\mnormal)
    = &\frac12\log\bigl(\det(2\pi e\cov{\isoreal{\mnormal}})\bigr)\\
    = &\frac12\log\bigl(\det(2\pi e\cov{\isoreal{\compmatrix}\isoreal{\smnormal}})\bigr)\\
    = &\frac12\log\bigl(\det(2\pi e\isoreal{\compmatrix}\cov{\isoreal{\smnormal}}
                       \isoreal{\compmatrix}^\trans)\bigr)\\
    = &\frac12\log\bigl(\det(\pi e\isoreal{\compmatrix}\isoreal{\compmatrix}^\trans)\bigr)
                        +\frac12\log\bigl(\det(2\cov{\isoreal{\smnormal}})\bigr)\\
    = &\frac12\log\bigl((\pi e)^{2\mixnumber}\det(\isoreal{(\compmatrix\compmatrix^\htrans)})\bigr)
                          +\frac12\log\bigl(\prod_{\dindex=1}^{\mixnumber}(1-\speccomp_{\dindex}^2)\bigr)\\
    = &\frac12\log\bigl((\pi e)^{2\mixnumber}\det(\isoreal{\cov{\mnormal}})\bigr)
                          +\frac12\sum_{\dindex=1}^{\mixnumber}\log(1-\speccomp_{\dindex}^2)\\
    = &\frac12\log\bigl((\pi e)^{2\mixnumber}\det(\cov{\mnormal})^2\bigr)
                          +\frac12\sum_{\dindex=1}^{\mixnumber}\log(1-\speccomp_{\dindex}^2)\\
    = &\log\bigl(\det(\pi e\cov{\mnormal})\bigr)
                          +\frac12\sum_{\dindex=1}^{\mixnumber}\log(1-\speccomp_{\dindex}^2)
 \end{split}
 \end{equation}
  by the properties of Lemma~\ref{lem:matrix_properties}.
\end{proof}

Since the summation term on the right of Eq.~(\ref{eqn:entropy_normal}) is always nonpositive
and the entropy of real r.vc.s with the given covariance is maximized for Gaussian r.vc.s \cite{Cover:1991}, it may be seen that the entropy of complex r.vc.s with the given covariance
is maximized for a narrow sense complex normal r.vc. \cite{Neeser:1993}, 
i.e., for a complex normal r.vc. with zero pseudo-covariance.
Theorem~\ref{thm:normal_rep} allows also an easy derivation
of the c.f. of a complex normal r.vc. \cite{Picinbono:1996,Vakhania:1996}.

\begin{seur}\label{thm:normal_cf}
  The c.f. of a complex normal r.vc. $\mnormal$ is given by
  \begin{equation}
  \begin{split}
    \cf_{\mnormal}(\compvector)
     =\exp\bigl(& -\frac14\compvector^\htrans\cov{\mnormal}\compvector
             -\frac14\real{\compvector^\htrans\pcov{\mnormal}\cconj{\compvector}}\\
                & +\jmath\real{\compvector^\htrans\E{\mnormal}{\mnormal}}\bigr)\\
     =\exp\bigl(& -\frac14\real{\inner{\compvector}
                             {\cov{\mnormal}\compvector
                             +\pcov{\mnormal}\cconj{\compvector}}}\\
                &+\jmath\real{\inner{\compvector}{\E{\mnormal}{\mnormal}}}\bigr).
  \end{split}
  \end{equation}
\end{seur}
\begin{proof}
  By Theorem~\ref{thm:normal_rep},
  $\mnormal=\compmatrix(\realpart{\smnormal}+\jmath\imagpart{\smnormal})+\mean$.
  Let $\compvector=\realpart{\compvector}+\jmath\imagpart{\compvector}\in\mathbb{C}^\mixnumber$,
  and $\smnormal=\realpart{\smnormal}+\jmath\imagpart{\smnormal}$.
  Now
  \begin{equation}
  \begin{split}
    \cf_{\smnormal}(\compvector)
    = & \cf_{\isoreal{\smnormal}}(\isoreal{\compvector})
      = \exp\bigl(-\frac12(\isoreal{\compvector}^\trans\cov{\isoreal{\smnormal}}
        \isoreal{\compvector})\bigr)\\
    = & \exp\bigl(-\frac14(\realpart{\compvector}^\trans(\identmatrix_{\mixnumber}+\diag{\specop})\realpart{\compvector}\\
      & +\imagpart{\compvector}^\trans(\identmatrix_{\mixnumber}-\diag{\specop})\imagpart{\compvector})\bigr)\\
    = & \exp\bigl(-\frac14(\realpart{\compvector}^\trans\realpart{\compvector}
                           +\imagpart{\compvector}^\trans\imagpart{\compvector}
                           +\realpart{\compvector}^\trans\diag{\specop}\realpart{\compvector}\\
      & -\imagpart{\compvector}^\trans\diag{\specop}\imagpart{\compvector})\bigr)\\
    = & \exp\bigl(-\frac14(\compvector^\htrans\compvector
                           +\real{\compvector^\trans\diag{\specop}\compvector})\bigr),
  \end{split}                         
  \end{equation}
  and by Eq.~(\ref{eqn:htrans})
  \begin{equation}
  \begin{split}
    \cf_{\mnormal}(\compvector)
    = & \cf_{\compmatrix\smnormal+\mean}(\compvector)
    =   \cf_{\compmatrix\smnormal}(\compvector)
        \exp\bigl(\jmath\real{\inner{\compvector}{\mean}}\bigr)\\
    = & \cf_{\smnormal}(\compmatrix^\htrans\compvector)
        \exp\bigl(\jmath\real{\inner{\compvector}{\mean}}\bigr)\\
    = & \exp\bigl(-\frac14(\compvector^\htrans\compmatrix\compmatrix^\htrans\compvector
                  +\real{\compvector^\trans\cconj{\compmatrix}\diag{\specop}
                         \compmatrix^\htrans\compvector})\bigr)\\
      & \exp\bigl(\jmath\real{\compvector^\htrans\mean}\bigr)\\       
    = & \exp\bigl(-\frac14(\compvector^\htrans\compmatrix\compmatrix^\htrans\compvector
                  +\real{\compvector^\htrans\compmatrix\diag{\specop}
                         \compmatrix^\trans\cconj{\compvector}})\\
      & +\jmath\real{\compvector^\htrans\mean}\bigr).
  \end{split}
  \end{equation}
\end{proof}

Corollary~\ref{thm:normal_cf} shows in particular 
that the \emph{second characteristic function} $\scf_{\mixture}\triangleq\log\cf_{\mixture}$
of a complex r.vc. $\mixture$ is a second-order wide sense polynomial in
variables $(\compvector,\cconj{\compvector})$.
Theorem~\ref{thm:normal_rep} can be also used to derive the density function
of a complex normal r.vc. However, unlike the c.f., the density function of a wide sense
normal r.vc. does not appear to have a simple form. See \cite{Picinbono:1996} for 
expressions for the density function in terms of the covariance and the pseudo-covariance matrices.
The following example essentially shows that in some cases the distribution of
a standard complex normal r.vc. is invariant to orthogonal transformations.

\begin{esim}\label{ex:nonident}
  Let the components of $\mnormal$ be uncorrelated complex normal r.v.s with
  the same circularity coefficient $\speccomp$. Now for a diagonal matrix $\diagmatrix$ the r.vc.
  $\diagmatrix\mnormal$ is standard complex normal with the circularity spectrum 
  $(\speccomp\ \cdots\ \speccomp)^\trans$, 
  and for any (real-valued) orthonormal matrix $\orthmatrix$,
  $\cov{\orthmatrix\diagmatrix\mnormal}
   =\orthmatrix\cov{\diagmatrix\mnormal}\orthmatrix^\htrans
   =\orthmatrix\identmatrix_{\mixnumber}\orthmatrix^\trans=\identmatrix_{\mixnumber}$ and
   $\pcov{\orthmatrix\diagmatrix\mnormal}
   =\orthmatrix\pcov{\diagmatrix\mnormal}\orthmatrix^\trans
   =\orthmatrix(\speccomp\identmatrix_{\mixnumber})\orthmatrix^\trans=\speccomp\identmatrix_{\mixnumber}$.
   Therefore, the r.vc. $\orthmatrix\diagmatrix\mnormal$ is also standard complex normal.
\end{esim}

\subsection{Darmois-Skitovich theorem for complex random variables}\label{ssc:cDSt}

One of the main characterization theorems for real r.v.s is the well-known
Darmois-Skitovich theorem (see \cite{Kagan:1973}). The theorem is fundamental
for proving the identifiability of real ICA models \cite{Comon:1994,Eriksson:2004}.
Here we extend the theorem to complex r.v.s. 

The proofs of the complex Darmois-Skitovich theorem and the proof of a closely related characterization
theorem (Theorem~\ref{thm:main_characterization} in Section~\ref{thm:main_characterization}) are both based on a complex
functional equation (Lemma~\ref{thm:complexFE} in Appendix~\ref{sec:proofDS}).
The functional equation is an extension of the corresponding
equation for real variables (see, e.g., Lemma 1.5.1 in \cite{Kagan:1973}) to complex variables.
Using the mapping (\ref{eqn:isomorphism}) Lemma~\ref{thm:complexFE} may be easily seen
to be a direct consequence
of the real multivariate theorem \cite{Ghurye:1962} 
(see also \cite{Kagan:1973,Mathai:1977}). 
A direct proof is given in Appendix~\ref{sec:proofDS} for the sake of completeness.

The complex extension of Darmois-Skitovich theorem has exactly the same form as the
real theorem with the \emph{wide sense} complex normal r.v.s taking the role
of real normal r.v.s. Hence, this theorem is an example where
the analogy \cite{Goodman:1963} between theories of
narrow sense complex normal r.v.s and real normal r.v.s is broken.

\begin{lause}[Complex Darmois-Skitovich]\label{thm:complexDS}
  Let $\sourcecomp_\dindex,\ldots,\sourcecomp_\altsounumber$ be mutually independent 
  complex r.v.s. If the linear forms (the r.v.s)
  \begin{equation}
    \mixturecomp_1=\sum_{\dindex=1}^\altsounumber\mixingel_\dindex\sourcecomp_\dindex
      \text{ and }
    \mixturecomp_2=\sum_{\dindex=1}^\altsounumber\altmixingel_\dindex\sourcecomp_\dindex,
  \end{equation}
  where $\mixingel_\dindex,\altmixingel_\dindex\in\mathbb{C}$, $\dindex=1,\ldots,\altsounumber$,
  are independent, then r.v.s
  $\sourcecomp_\dindex$ for which $\mixingel_\dindex\altmixingel_\dindex\neq 0$
  are complex normal.
\end{lause}
\begin{proofwl}{Sketch of the proof}
 The complete proof is given in Appendix~\ref{sec:proofDS} and it
 follows the proof of the real-valued Darmois-Skitovich theorem
 (see \cite{Kagan:1973}) with appropriate extensions to complex field. The idea
 is to consider two forms of the logarithm of the 
 joint c.f. of $\mixturecomp_1$ and $\mixturecomp_2$ following from independence.
 This functional equation is only satisfied for wide sense polynomials showing that 
 the r.v. $\mixturecomp_1$ is complex normal. This is only possible if
 r.v.s $\sourcecomp_\dindex$ are complex normal.
\end{proofwl}

Although narrow sense complex normal r.v.s had to be admitted to 
the complex Darmois-Skitovich theorem,
it may still appear in the view of Corollary~\ref{thm:unique_swhite}
that complex normal r.v.s appearing in the theorem can not be completely
arbitrary. That is, it may appear that some of the circularity coefficients 
of normal r.v.s should be equal. It is true if $\altsounumber=2$.
However, it is not generally true as
it is shown in the next example.

\begin{esim}
  Let $\smnormal_1=(\normal_1,\normal_2,\normal_3)^\trans$ be standard complex normal r.vc. with
  the circularity spectrum $\spectrum{\smnormal_1}=(\frac13,\frac15,\frac18)^\trans$. Then
  $\smnormal_2=\frac1{5\sqrt{2}}
  \bigl(\begin{smallmatrix} 3 & 5 & 4\\ 3 & -5 & 4\end{smallmatrix}\bigr)\smnormal_1$
  is also standard complex normal r.vc. with the circularity spectrum  
  $\spectrum{\smnormal_2}=(\frac15,\frac15)^\trans$. Thus marginals of 
  $\smnormal_2$ are independent, and the Darmois-Skitovich theorem applies.
  However, the circularity spectrum of $\smnormal_1$ is distinct. Notice also that by
  Example~\ref{ex:nonident},
  the r.vc. obtained from $\smnormal_2$ by multiplying with any orthogonal
  matrix is also standard complex normal r.vc. with the same circularity spectrum. 
\end{esim}

\section{Complex {ICA} Models}\label{sec:complex_ica}

In this section, we show that complex ICA is actually a well-defined concept, and
we establish theoretical conditions similar to the real-valued case \cite{Eriksson:2004}.
In Section~\ref{ssc:ICAdefs} the main definitions along with some illustrative
examples are given. Also a crucial characterization theorem  giving a connection
between vector coefficients and complex normal r.v.s is proved.
Finally, in sections \ref{ssc:seprability}, \ref{ssc:identifiability}, and
\ref{ssc:uniqueness} the conditions for separability, identifiability,
and uniqueness of complex ICA models, respectively, are derived.

\subsection{Definitions and problem statement}\label{ssc:ICAdefs}

A general linear instantaneous complex-valued ICA
model may be described by the equation
\begin{equation}\label{eqn:model}
  \mixture=\mixing\source,
\end{equation}
where 
$(\sourcecomp_1,\ldots,\sourcecomp_{\sounumber})^{\trans}=\source$ 
are unknown complex-valued independent non-degenerate r.v.s, 
i.e., \emph{sources},
$\mixing$ is a complex constant $\mixnumber\times\sounumber$ unknown mixing matrix,
$\mixnumber\geq2$, and
$\mixture=(\mixturecomp_1,\ldots,\mixturecomp_\mixnumber)^{\trans}$ are \emph{mixtures}, i.e.,
the observed complex r.vc. (sensor array output). 
The couple
$(\mixing,\source)$ is called a \emph{representation} of r.vc. $\mixture$.
If no column in the mixing matrix $\mixing$ is \emph{collinear} with another column in the matrix,
i.e., all columns are pairwise linearly independent,
the representation is called \emph{reduced}. All representations 
are assumed to be reduced throughout this paper. Furthermore, a reduced representation
for the r.vc. $\mixture$ in the model~(\ref{eqn:model}) is called \emph{proper},
if it satisfies all the assumptions made about the model.

The model of Eq.~(\ref{eqn:model}) is defined to be
\begin{enumerate}[(i)] 
  \item \emph{identifiable},
      or the mixing matrix is (essentially) unique, if in every proper representations
      $(\mixing,\source)$ and  $(\altmixing,\altsource)$ of $\mixture$, 
      every column of complex matrix $\mixing$
      is collinear with a column of complex matrix $\altmixing$ and vice versa,
  \item \emph{unique} if the
      model is identifiable and furthermore the source r.vc.s $\source$
      and $\altsource$ in different proper representations 
      have the same distribution for some permutation
      up to changes of location and complex scale, and
  \item \emph{separable}, if for every complex matrix $\solution$ 
      such that $\solution\mixture$ has $\sounumber$ independent components, we have
      $\diagmatrix\permmatrix\source=\solution\mixture$
      for some diagonal matrix $\diagmatrix$ with nonzero diagonals
      and permutation matrix $\permmatrix$. Moreover, such a matrix $\solution$
      has to always exist.
\end{enumerate}

It is completely possible for the model~(\ref{eqn:model})
to be identifiable but not unique nor separable as it is
shown in the next example. 

\begin{esim}\label{ex:nonunique}
 As an example of a model which is identifiable
 but is not separable nor unique, 
 consider independent non-normal r.v.s $\sourcecomp_k$, $k=1,\ldots 4$. Let
 $\snormal_1$, $\snormal_2$, and 
 $\snormal_3$
 be independent standard normal r.v.s
 with the same circularity coefficient. 
 Then also 
 r.v.s $\snormal_1+\snormal_2$ and $\snormal_1-\snormal_2$ are independent. Now
 \begin{equation}
 \begin{split}
\begin{pmatrix}
   \sourcecomp_1+\sourcecomp_3+\sourcecomp_4+\snormal_1+\snormal_2\\
   \sourcecomp_2+\sourcecomp_3-\sourcecomp_4+\snormal_1-\snormal_2
  \end{pmatrix}
 =  \begin{pmatrix}
   1 & 0 & 1 &  1\\ 
   0 & 1 & 1 & -1
  \end{pmatrix} 
 \begin{pmatrix}
    \sourcecomp_1\\
    \sourcecomp_2\\
    \sourcecomp_3+\snormal_1\\
    \sourcecomp_4+\snormal_2
  \end{pmatrix} \\
  = 
  \begin{pmatrix}
   1 & 0 & 1 & 1\\ 
   0 & 1 & 1 & -1
  \end{pmatrix}
 \begin{pmatrix}
    \sourcecomp_1+\snormal_1+\snormal_2\\
    \sourcecomp_2+\snormal_1-\snormal_2\\
    \sourcecomp_3\\
    \sourcecomp_4
  \end{pmatrix},
 \end{split}
 \end{equation}
 which shows that the corresponding model can not be unique. However, it is identifiable.
 R.v.s of the form $\sourcecomp+\normal$, where $\normal$ is a normal r.v. independent of $\sourcecomp$, are said to have a 
 \emph{normal component}.
\end{esim}

It follows from the reduction assumption that
the number of columns, i.e., the number of sources or the \emph{model order}, 
is the same in every proper representation of $\mixture$ in identifiable models. 
If $\solution$ is a separating matrix, then linear manifolds of $\diagmatrix\permmatrix$ and $\solution$ must coincide, and therefore 
$\mixnumber\geq\rank{\solution}=\rank{\diagmatrix\permmatrix}=\sounumber$,
i.e., there has to be at least as many mixtures as sources in a separable model.
This fact also emphasizes that identifiability of the model~(\ref{eqn:model}) depends
also on the linear operator structure, and since the linear operators defined on 
$\mathbb{R}^{2\mixnumber}$ and $\mathbb{C}^{\mixnumber}$ are not isomorphic,
one can not simply consider real-valued model with twice the observation dimension
when studying the complex ICA model~(\ref{eqn:model}). This is illustrated in
the following example.

\begin{esim}
By simply considering real-valued models with twice the dimension, it may actually
seem that the complex separation is possible only under very strict conditions.
Indeed, let $\altsourcecomp_{\dindex}$, $k=1\ldots,4$, be independent real-valued 
r.v.s, and let
$\mixing_1$, $\mixing_2$, $\altmixing_1$, and $\altmixing_2$ be $2\times2$ nonsingular
real matrices. Define $\source_1=\mixing_1(\altsourcecomp_1\,\,\altsourcecomp_2)^\trans$ and
$\source_2=\mixing_2(\altsourcecomp_3\,\,\altsourcecomp_4)^\trans$. Now
$\source_1$ and $\source_2$ are independent, but so are also $\soluvec_1$ and
$\soluvec_2$,
\begin{equation}\label{eqn:badmixture}
    \begin{pmatrix}
      \soluvec_1\\
      \soluvec_2         
    \end{pmatrix}
    =
    \begin{pmatrix}
      \altmixing_1 & \multizero_{2\times2}  \\
      \multizero_{2\times2}    & \altmixing_2
    \end{pmatrix}
    \permmatrix
    \begin{pmatrix}
      \mixing_1^{-1} & \multizero_{2\times2}   \\
      \multizero_{2\times2}    & \mixing_2^{-1}
    \end{pmatrix}
    \begin{pmatrix}
      \source_1\\
      \source_2         
    \end{pmatrix},
\end{equation}
for any permutation matrix $\permmatrix$. However, $\soluvec_1$ and
$\soluvec_2$ are \emph{mixtures} of $\source_1$ and $\source_2$ for many
permutations $\permmatrix$.
\end{esim}

The previous example is easily generalized to the ICA models that have 
\emph{multidimensional} independent sources, i.e., one is looking for independent multidimensional subspaces.
The example shows that such models can not 
be identified or separated without additional constraints on the internal
dependency structure of the sources or the allowed mixing matrices. 

Since linear operators in complex and real spaces are not isomorphic,
the classes of separable source r.v.s are not the same.
That is, some source r.v.s considered in complex mixtures can be separated
although their real-valued representations in real mixtures can not.
This is shown in the next example.

\begin{esim}\label{ex:normalseparable}
  Let $\snormal_1,\ldots,\snormal_{2\sounumber}$ be independent standard zero mean
  unit variance real Gaussian r.v.s. Define
  \begin{equation}
  \begin{split}
  \smnormal=\bigl(
   \frac1{\sqrt{\sounumber+1}}(\sqrt{\sounumber}\snormal_1+\jmath\snormal_{\sounumber+1})&,
   \frac1{\sqrt{\sounumber}}(\sqrt{\sounumber-1}\snormal_2+\jmath\snormal_{\sounumber+2}),\\
   \ldots &,\frac1{\sqrt{2}}(\snormal_\sounumber+\jmath\snormal_{2\sounumber})\bigr).
   \end{split}
   \end{equation}
   Now it is easily seen that $\smnormal$ is a standard normal r.vc. with the distinct
   circularity spectrum $\spectrum{\smnormal}=
   (\frac{\sounumber-1}{\sounumber+1},\frac{\sounumber-2}{\sounumber},\ldots,0)^\trans$.
   If $\isoreal{\smnormal}$ is taken as the source r.vc. in the real-valued ICA model,
   i.e., $\soluvec=\altmixing\isoreal{\smnormal}$ and  $\altmixing$ is a 
   $2\mixnumber\times2\sounumber$ real-valued matrix, $\mixnumber\geq\sounumber$,
   the model is not separable \cite{Eriksson:2004}.
   However, the complex model involving
   $\smnormal$ itself, i.e., 
   $\mixture=\mixing\smnormal$ and  $\mixing$ is a 
   $\mixnumber\times\sounumber$ complex-valued matrix,
   is separable by Corollary~\ref{thm:unique_swhite}.
\end{esim}

The following characterization theorem 
is the base of the identifiablility and uniqueness theorems.
It is an extension of a real theorem
\cite[Theorem 10.3.1]{Kagan:1973} to the complex case. 
The idea of the proof is similar to the proof of Darmois-Skitovich theorem, and the
proof given follows loosely that of the real counterpart with appropriate complex extensions.

\begin{lause}\label{thm:main_characterization}
  Let $(\mixing,\source)$ and $(\altmixing,\altsource)$
  be two reduced representations of a $\mixnumber$-dimensional complex r.vc. $\mixture$,
  where $\mixing$ and $\altmixing$ are constant complex matrices of dimensions
  $\mixnumber\times\sounumber$ and $\mixnumber\times\altsounumber$, respectively, and
  $\source=(\sourcecomp_1,\ldots,\sourcecomp_\sounumber)^\trans$ and 
  $\altsource=(\altsourcecomp_1,\ldots,\altsourcecomp_\altsounumber)^\trans$ are 
  complex r.vc.s with independent 
  components. Then the following properties hold.
  \begin{enumerate}[(i)]
    \item If the $\dindex$th column of $\mixing$ is not collinear with
          any column of $\altmixing$, then the r.v. $\sourcecomp_\dindex$
          is complex normal. \label{thm:mc_part1}
    \item If the $\dindex$th column of $\mixing$ is collinear with
          the $\dindextwo$th column of $\altmixing$, then the
          logarithms of the c.f.s of
          r.v.s $\sourcecomp_\dindex$ and $\altsourcecomp_\dindextwo$ differ
          by a wide sense polynomial in a neighborhood of the origin.\label{thm:mc_part2}
  \end{enumerate}
\end{lause}
\begin{proof}
\begin{enumerate}[(i)]
  \item By Lemma~\ref{thm:twodimmatrices} (see Appendix~\ref{sec:prooflemmas}), there exists
        a $2\times\mixnumber$ matrix $\compmatrix$ such that the $\dindex$th column of
        $\compmatrixtwo_1=\compmatrix\mixing$ is not collinear 
        with any other column of $\compmatrixtwo_1$,
        or with any column of  $\compmatrixtwo_2=\compmatrix\altmixing$. Then
        $\compmatrix\mixture=\compmatrixtwo_1\source=\compmatrixtwo_2\altsource$,
        and applying
        Lemma~\ref{thm:twodim_characterization}(\ref{thm:tc_part1}) (see Appendix~\ref{sec:prooflemmas}) it is seen that 
        the r.v. $\sourcecomp_k$
        is complex normal.
  \item By definitions the $\dindex$th column of $\mixing$, say $\mixingcol$, is collinear only with
        the $\dindextwo$th column of $\altmixing$, say $\altmixingcol$. Therefore 
        by Lemma~\ref{thm:twodimmatrices} (see Appendix~\ref{sec:prooflemmas}), there exists
        a $2\times\mixnumber$ matrix $\compmatrix$ such that the $\dindex$th column of
        $\compmatrixtwo_1=\compmatrix\mixing$ is not collinear 
        with any other columns of $\compmatrixtwo_1$,
        or with any column of  $\compmatrixtwo_2=\compmatrix\altmixing$ 
        except possibly the $\dindextwo$th. Furthermore, since
        $\compmatrix\mixingcol=\compmatrix(\cconst\altmixingcol)=\cconst(\compmatrix\altmixingcol)$ 
        for some $\cconst\in\mathbb{C}$,
        it is seen  that $(\compmatrixtwo_1,\source)$ and $(\compmatrixtwo_2,\altsource)$ are
        reduced representations of $\compmatrix\mixture$ such that
        Lemma~\ref{thm:twodim_characterization}(\ref{thm:tc_part2}) gives the claim.
\end{enumerate}
\end{proof}

\subsection{Separability}\label{ssc:seprability}

ICA is commonly used as a Blind Source Separation-method, where
the problem is to \emph{extract} the original signals from the observed
linear mixture.
Therefore, separability of the ICA model is an important issue.
The separability theorem for the complex ICA model below may be surprising, 
since it allows also separation of some complex
normal mixtures.

\begin{lause}[Separability]\label{thm:separability}
 The model of Eq. (\ref{eqn:model}) is separable if and only if
 the complex mixing matrix $\mixing$ is of full column rank and there are no two complex
 normal source r.v.s with the same circularity coefficient. 
\end{lause}
\begin{proof}
  Suppose the model is separable.
  Since $\sounumber=\rank{\solution\mixing}\leq\rank{\mixing}\leq\sounumber$,
  the mixing matrix $\mixing$ is of full column rank $\sounumber$.
  If there were two complex normal source r.v.s with the same circularity coefficient,
  by Example~\ref{ex:nonident} in Section~\ref{ssc:cGrvc},
  there would exist matrices that produce $\sounumber$ independent components
  but which are not diagonal matrices for any permutation of the columns.
  
  To the other direction, suppose the mixing matrix $\mixing$ is of full column rank and 
  there are no two complex
  normal source r.v.s with the same circularity coefficient. 
  Now $\mixing^\pinv$, where the superscript $\pinv$ denotes the Moore-Penrose
  generalized inverse \cite{Horn:1985}, is a separating matrix.
  Suppose $\solution$ is a matrix 
  such that $\solution\mixture$ has $\sounumber$ independent components.
  If $\solution\mixing$ is not of the form $\diagmatrix\permmatrix$, then
  there exist at least two columns such that they both contain at least two nonzero elements.
  By Lemma~\ref{thm:degen_independence} (see Appendix~\ref{sec:prooflemmas})
  there can not exist only one such column since the sources are nondegenerate.
  Assume without loss of generality that the first $\dindextwo$ columns
  $\altmixingcol_{\dindex}$, $\dindex=1,\ldots,\dindextwo\leq\sounumber$, of 
  $\solution\mixing$ are columns with at least two nonzero elements, 
  and denote the corresponding matrix of rank $\dindextwo$ by
  $\altmixing=
   (\altmixingcol_1\ \cdots\ \altmixingcol_\dindextwo)$.
  By Theorem~\ref{thm:complexDS} the r.v. $\sourcecomp_\dindex$ corresponding to 
  the column $\altmixingcol_{\dindex}$,
  $\dindex=1,\ldots,\dindextwo$, is complex normal, and we assume, without loss of
  generality, that  the r.vc.
  $\smnormal_1=
  (\sourcecomp_1\ \cdots\ \sourcecomp_\dindextwo)^\trans$
  is standard complex normal.
  By Theorem~\ref{thm:cramer} (see Appendix~\ref{sec:proofDS}) all components of  
  $\mnormal_2=\altmixing\smnormal_1$
  are complex normal, and by Lemma~\ref{thm:sum_independence} (see Appendix~\ref{sec:prooflemmas}) all components of $\mnormal_2$ are
  independent. Choose any $\dindextwo$ rows of $\altmixing$ such
  that the corresponding submatrix $\hat{\altmixing}$ is of rank $\dindextwo$,
  and $\hat{\altmixing}$ contains a row with two nonzero elements. Since
  $\hat{\altmixing}$ is not diagonal for any permutation by construction,
  $\smnormal_1$ is standard, and $\mnormal_2$ has independent components,
  it follows from Corollary~\ref{thm:unique_swhite} that $\smnormal_1$ can not have
  a distinct circularity spectrum, which is a contradiction. Therefore, $\solution\mixing$ is of the form 
  $\diagmatrix\permmatrix$, and the model is separable.
\end{proof}

\begin{huom}
  If the source $\source$ has finite second order statistics and the circularity spectrum
  $\spectrum{\source}$ is distinct, then 
  the separation can be achieved by simply performing the strong-uncorrelating transform by Corollary~\ref{thm:unique_swhite}.
  In this case, there is no additional restrictions on the distribution of the source r.v.s, and therefore 
  some \emph{normal} r.v.s can be also separated. An
  example of such a mixture is seen in Example~\ref{ex:normalseparable}.
\end{huom}

\subsection{Identifiability}\label{ssc:identifiability}

Identifiability considers reconstruction of the mixing matrix. This
is useful in some problems, where the immediate
interest may not be in the sources themselves but in how they were mixed
(e.g., channel matrix in MIMO communications).

\begin{lause}[Identifiability]\label{thm:identifiability}
 The model of eq.~(\ref{eqn:model}) is identifiable, if
 \begin{enumerate}[(i)]
    \item no source r.v. is complex normal, or \label{thm:identifiability_case1}
    \item $\mixing$ is of full column rank and there are no two complex
          normal source r.v.s with the same circularity coefficient. \label{thm:identifiability_case2}
 \end{enumerate}
\end{lause}
\begin{proof}
  \begin{enumerate}[(i)]
    \item Since there are no complex normal r.v.s, by 
        Theorem~\ref{thm:main_characterization}(\ref{thm:mc_part1}),
        every column has to be collinear with exactly a column in another
        proper representation, i.e., the model is identifiable. 
    \item Let$(\mixing,\source)$ and  $(\altmixing,\altsource)$ be proper representations
        of $\mixture$. Since the model is separable 
        by Theorem~\ref{thm:separability}
        and $\mixing^\pinv$ is a separating matrix, 
        $\mixing^\pinv\altmixing=\permmatrix\diagmatrix$ for a permutation matrix
        $\permmatrix$ and a diagonal matrix $\diagmatrix$. By the uniqueness of the 
        generalized inverse, it follows $\mixing\permmatrix\diagmatrix=\altmixing$.
 \end{enumerate}
\end{proof}

There is a striking contrast between the two cases in Theorem~\ref{thm:identifiability}.
Namely, if there are more sources than mixtures not a single normal r.v. is allowed
whereas in the other case all source r.v.s can be normal. The following example
shows the reason why we can not allow a single normal r.v. for identifiability when
there are more sources than sensors.

\begin{esim}\label{ex:anormnonident}
 Consider independent non-normal r.v.s $\sourcecomp_1, \sourcecomp_2$, and
 standard normal r.v.s $\snormal_1$ and $\snormal_2$
 with the same circularity coefficient. 
 Now
 \begin{equation}
 \begin{split}
 \mixture =
  \begin{pmatrix}
   \sourcecomp_1+\sourcecomp_2+2\snormal_1\\
   \sourcecomp_1+2\snormal_2
  \end{pmatrix} = &
  \begin{pmatrix}
   1 & 1 & 0 \\ 
   1 & 0 & 1 
  \end{pmatrix}
  \begin{pmatrix}
    \sourcecomp_1\\
    \sourcecomp_2+2\snormal_1\\
    2\snormal_2
  \end{pmatrix}\\
  = & 
  \begin{pmatrix}
   1 & 1 & 1 \\ 
   1 & 0 & -1 
  \end{pmatrix}
  \begin{pmatrix}
    \sourcecomp_1+\snormal_1+\snormal_2\\
    \sourcecomp_2\\
    \snormal_1-\snormal_2
  \end{pmatrix},
 \end{split}
 \end{equation}
 and the last column shows that the model is not identifiable. 
\end{esim}

It is evident from the previous example and from the separation theorem that
another  identifiability condition could be formulated by essentially allowing
a single normal r.v. and not allowing other source r.v.s to have normal
components with the same circularity coefficient. 
However, this condition is unnecessarily complicated.
Therefore, it is not stated in a formal manner.

\subsection{Uniqueness}\label{ssc:uniqueness}

Uniqueness considers the case where one is interested not only in the mixing matrix
but also in the distribution of the sources.

\begin{lause}[Uniqueness]\label{thm:uniqueness}
 The model of Eq.~(\ref{eqn:model}) is unique if either of the following
 properties hold.
  \begin{enumerate}[(i)]
    \item The model is separable.\label{thm:uniqueness_case1}
    \item All c.f.s  of source r.v.s are analytic (or all c.f.s are non-vanishing), 
          and none of the c.f.s  has an exponential factor with 
          a wide sense polynomial of degree
          at least two,
          i.e., no source r.v. has the c.f. $\cf$ such that
          $\cf(\cocomp)=\cf_1(\cocomp)\exp(\poly(\cocomp,\cconj{\cocomp}))$ for 
          a c.f. $\cf_1(\cocomp)$ and for some
          wide sense polynomial 
          $\poly(\cocomp,\cconj{\cocomp})$ of degree at least two. \label{thm:uniqueness_case2}
 \end{enumerate}
\end{lause}
\begin{proof}
\begin{enumerate}[(i)]
  \item Let$(\mixing,\source)$ and  $(\altmixing,\altsource)$ be proper representations
        of $\mixture$. By Theorem~\ref{thm:identifiability}(\ref{thm:identifiability_case2})
        the model is identifiable, and therefore $\mixing\permmatrix\diagmatrix=\altmixing$
        for a permutation matrix $\permmatrix$ and a diagonal matrix $\diagmatrix$.
        Now 
        $\source=\mixing^\pinv\mixture=\mixing^\pinv\altmixing\altsource=\permmatrix\diagmatrix\altsource$.
  \item There can not be any complex normal r.v.s, and therefore the model is identifiable by  
        Theorem~\ref{thm:identifiability}(\ref{thm:identifiability_case1}).  
        Now the logarithms of the c.f.s of the source variables
        in two proper representations differ by a wide sense polynomial by
        Theorem~\ref{thm:main_characterization}(\ref{thm:mc_part2}). However,
        by the assumption this wide sense polynomial can be at most of degree $1$, i.e.,
        the source variables have the same distribution up to changes of location and complex scale.
\end{enumerate}
\end{proof}

A nonunique but identifiable mixture was described in Example~\ref{ex:nonunique}.
By slightly restricting the allowed mixing matrices, it is possible in the real case 
to obtain more classes of unique models \cite{Eriksson:2004}. Further work is needed
to determine if those
theorems can be extended to the complex case.

\section{Conclusion}

In this paper conditions for
separability, identifiablity, and uniqueness of complex-valued linear
ICA models are established. 
Both  
circular and noncircular complex random vectors are covered by the results.
So far these conditions have been known for real random vectors only.
The conditions for 
identifiablity, and uniqueness are sufficient and 
the separability condition is also found to be necessary.
In order to show these results, a proof of  complex extension of the Darmois-Skitovich 
Theorem is constructed. Some second-order properties and 
characterizations of linear forms of complex random vectors are 
reviewed and new results found in the process of proving the theorem.
As a by-product of establishing the conditions, a theorem on
differential entropy for complex normal random vectors is proved and a slightly surprising result
about separating complex Gaussian sources is found.

\section*{Acknowledgment}

The authors wish to thank the anonymous reviewers for their valuable comments and suggestions.

\appendices

\section{Proof of Lemma~\ref{thm:cc_properties}}\label{sec:proofcclemma}

\begin{proofwl}{Proof of Lemma~\ref{thm:cc_properties}}
  By Theorem~\ref{thm:compdef} there exist nonzero constants $\cconstthree,\cconstfour\in\mathbb{C}$
  such that r.v.s $\sourcecomp=\cconstthree\mixturecomp$ and $\altsourcecomp=\cconstfour\solucomp$
  are strongly uncorrelated.
  \begin{enumerate}[(i)]
    \item Since $\cov{\realpart{\sourcecomp}}+\cov{\imagpart{\sourcecomp}}=1$,
          $0\leq\spectrum{\mixturecomp}=\spectrum{\sourcecomp}
          =\cov{\realpart{\sourcecomp}}-\cov{\imagpart{\sourcecomp}}
          =1-2\cov{\imagpart{\sourcecomp}}\leq1$.
          Also $\frac{\cconstthree}{\cconst}(\cconst\mixturecomp)=\sourcecomp$, and thus by uniqueness
          $\spectrum{\cconst\mixturecomp}=\spectrum{\sourcecomp}=\spectrum{\mixturecomp}$. Furthermore
          \begin{equation}
          \begin{split}
            \spectrum{\mixturecomp}= &\pcov{\sourcecomp} 
           =\frac{|\pcov{\sourcecomp}|}{\cov{\sourcecomp}}
           =\frac{|\pcov{\cconstthree\mixturecomp}|}{\cov{\cconstthree\mixturecomp}}\\
           = &\frac{|\cconstthree^2\pcov{\mixturecomp}|}{|\cconstthree|^2\cov{\mixturecomp}}
             =\frac{|\cconstthree^2||\pcov{\mixturecomp}|}{|\cconstthree|^2\cov{\mixturecomp}}
             =\frac{|\pcov{\mixturecomp}|}{\cov{\mixturecomp}}.
          \end{split}
          \end{equation}
    \item $\spectrum{\mixturecomp}=1-2\cov{\imagpart{\sourcecomp}}=1$ if and only if
          $\cov{\imagpart{\sourcecomp}}=0$.
    \item Suppose $\spectrum{\mixturecomp}\geq\spectrum{\solucomp}$. 
          Using the first part of the lemma for an r.v. 
          $\mixturecomp+\solucomp$, uncorrelateness, and the triangle inequality, we have
           \begin{equation}\label{eqn:coefficient_ieq}
           \begin{split}
              \spectrum{\mixturecomp+\solucomp} &
                =\frac{|\pcov{\mixturecomp+\solucomp}|}{\cov{\mixturecomp+\solucomp}}
                =\frac{|\pcov{\mixturecomp}+\pcov{\solucomp}|}{\cov{\mixturecomp}+\cov{\solucomp}}\\
              = &\frac{|\pcov{\frac1{\cconstthree}\sourcecomp}+\pcov{\frac1{\cconstfour}\altsourcecomp}|}
                      {\cov{\frac1{\cconstthree}\sourcecomp}+\cov{\frac1{\cconstfour}\altsourcecomp}}\\
              = &\frac{|\frac1{\cconstthree^2}\pcov{\sourcecomp}
                +\frac1{\cconstfour^2}\pcov{\altsourcecomp}|}
                      {\frac1{|\cconstthree|^2}+\frac1{|\cconstfour|^2}}\\
              = &\frac{|\frac1{\cconstthree^2}\spectrum{\mixturecomp}
                       +\frac1{\cconstfour^2}\spectrum{\solucomp}|}
                      {\frac1{|\cconstthree|^2}+\frac1{|\cconstfour|^2}}\\
              & \leq \frac{\frac1{|\cconstthree|^2}\spectrum{\mixturecomp}
                       +\frac1{|\cconstfour|^2}\spectrum{\solucomp}}
                      {\frac1{|\cconstthree|^2}+\frac1{|\cconstfour|^2}}\leq\spectrum{\mixturecomp},
           \end{split}
           \end{equation}
           which proves the inequality. 

           If both r.v.s $\mixturecomp$ and $\solucomp$ are second order circular, then clearly the equality holds
           in (\ref{eqn:coefficient_ieq}).
           Now suppose the condition for the equality holds in the noncircular case, and let
           $\speccomp=\spectrum{\mixturecomp}=\spectrum{\solucomp}$ and
           $\cargu=\Arg(\pcov{\mixturecomp})=\Arg(\pcov{\solucomp})$. Then
           \begin{equation}
           \begin{split}
              \spectrum{\mixturecomp+\solucomp} 
                & = \frac{|\pcov{\mixturecomp}+\pcov{\solucomp}|}{\cov{\mixturecomp}+\cov{\solucomp}}\\
                & = \frac{|\speccomp\cov{\mixturecomp}e^{\jmath\cargu}+\speccomp\cov{\solucomp}e^{\jmath\cargu}|}{\cov{\mixturecomp}+\cov{\solucomp}}\\
                & =  \frac{|\speccomp e^{\jmath\cargu}||\cov{\mixturecomp}+\cov{\solucomp}|}{\cov{\mixturecomp}+\cov{\solucomp}}
                  = \speccomp.%
           \end{split}
           \end{equation}
           To the other direction, the last inequality in (\ref{eqn:coefficient_ieq}) holds with the equality iff
           $\spectrum{\mixturecomp}=\spectrum{\solucomp}$. If now $\spectrum{\mixturecomp}\neq0$, 
           then the triangle inequality in (\ref{eqn:coefficient_ieq}) holds with the equality iff 
           \begin{equation}
              0 <\frac{\cconstfour^2}{\cconstthree^2}
                =\frac{\cconstfour^2\pcov{\sourcecomp}}{\cconstthree^2\pcov{\altsourcecomp}}
                =\frac{\pcov{\mixturecomp}}{\pcov{\solucomp}}.
           \end{equation}
           Hence  $\Arg(\pcov{\mixturecomp})=\Arg(\pcov{\solucomp})$ by the polar forms of $\pcov{\mixturecomp}$ and $\pcov{\solucomp}$.
  \end{enumerate}
\end{proofwl}

\section{Proof of the complex Darmois-Skitovich theorem and related theorems}\label{sec:proofDS}

The following theorem is a direct consequence of the multivariate version
of the real Marcinkiewicz theorem. The theorem shows
essentially that a complex normal r.v. is the only r.v. whose second c.f.
is a wide sense polynomial.

\begin{lause}[Complex Marcinkiewicz]\label{thm:marcinkiewicz}
 If in some neighborhood of zero the c.f. $\cf_{\mixturecomp}$ of a complex r.v. $\mixturecomp$
 admits the representation
 \begin{equation}\label{eqm:marc_rel}
    \cf_{\mixturecomp}(\cocomp)=\exp\bigl(\poly(\cocomp,\cconj{\cocomp})\bigr),
 \end{equation}
  where $\poly$ is a wide sense polynomial, then the r.v. $\mixturecomp$ is complex normal.
\end{lause}
\begin{proof}
  Fix $\cocomp_0\in\mathbb{C}$, and define a c.f. 
  $\cf_0(\recomp)\triangleq\cf_{\mixturecomp}(\recomp\cocomp_0)
   =\exp\bigl(\poly(t\cocomp_0,t\cconj{\cocomp_0})\bigr)$ 
  for $\recomp\in\mathbb{R}$. Then for some
  $\varepsilon>0$, $\log\cf_0(\recomp)$ is a polynomial in $\recomp$, $|\recomp|<\varepsilon$.
  Therefore, by a version of $\alpha$-decomposition theorem (see \cite[Theorem 7.4.2]{Linnik:1977})
  the relation is valid for all $\recomp$ and $\cf_0(\recomp)$ is normal. 
  Since $\cocomp_0$ is assumed to be arbitrary, it follows that the equation (\ref{eqm:marc_rel})
  is valid for all $\cocomp$. By the last property of Lemma~\ref{lem:matrix_properties},
  $\poly(\cocomp,\cconj{\cocomp})$ is a polynomial in $\isoreal{\cocomp}$,
  and the claim follows from
  the multivarite (bivariate) Marcinkiewicz's theorem
  (e.g., \cite[Theorem 3.4.3]{Cuppens:1975}).
\end{proof}

Also the well-known Cramer's theorem has a direct complex counterpart.

\begin{lause}[Complex Cramer]\label{thm:cramer}
  If $\sourcecomp_1$ and $\sourcecomp_2$ are independent r.v.s such that 
  $\sourcecomp_1+\sourcecomp_2$ is a complex normal r.v., then each of the r.v.s
  $\sourcecomp_1$ and $\sourcecomp_2$ is complex normal.
\end{lause}
\begin{proof}
 This is a direct corollary to the
 real multivariate Cramer's theorem (e.g., \cite[Theorem 6.3.2]{Linnik:1977}).
\end{proof}

\begin{lemma}\label{thm:complexFE}
  Consider the equation, assumed valid for $|\cocomp_1|,|\cocomp_2|<\varepsilon$,
  \begin{equation}\label{eqn:poly_characterization}
    \sum_{\dindex=1}^{\mixnumber}\scf_{\dindex}(\cocomp_1+\cconst_{\dindex}\cocomp_2)
    =\cfunc_1(\cocomp_1)+\cfunc_2(\cocomp_2),
  \end{equation}
  where $\scf_{\dindex}$, $\dindex=1,\ldots,\mixnumber$, $\cfunc_1$, and $\cfunc_2$ are
  continuous complex-valued functions of complex variables 
  and the nonzero complex numbers $\cconst_{\dindex}$,
  $\dindex=1,\ldots,\mixnumber$, are distinct. Then all the functions in  
  (\ref{eqn:poly_characterization}) are wide sense polynomials in $(\cocomp,\cconj{\cocomp})$ of
  degree not exceeding $\mixnumber$.
\end{lemma}
\begin{proof}
  Let $\cconsttwo_{\dindex}^{(1)}=(1-\frac{\cconst_{\dindex}}{\cconst_{\mixnumber}})\cconstfour_1$.
  Now, for small enough $\cconstfour_1$, we have
  \begin{equation}\label{eqn:poly_chara_subs1}
  \begin{split}
     \sum_{\dindex=1}^{\mixnumber}\scf_{\dindex}(\cocomp_1+\cconstfour_1
           +\cconst_{\dindex}(\cocomp_2-\frac{\cconstfour_1}{\cconst_{\mixnumber}})) 
    = & \sum_{\dindex=1}^{\mixnumber}\scf_{\dindex}(\cocomp_1
                                 +\cconsttwo_{\dindex}^{(1)}+\cconst_{\dindex}\cocomp_2)\\
    = & \cfunc_1(\cocomp_1+\cconstfour_1)+\cfunc_2(\cocomp_2-\frac{\cconstfour_1}{\cconst_{\mixnumber}})
  \end{split}
  \end{equation}
  by substituting  $(\cocomp_1+\cconstfour_1)$ for $\cocomp_1$ and 
  $(\cocomp_2-\frac{\cconstfour_1}{\cconst_{\mixnumber}})$ for $\cocomp_2$ 
  in (\ref{eqn:poly_characterization}).
  Subtracting (\ref{eqn:poly_characterization}) from (\ref{eqn:poly_chara_subs1}), we obtain
  \begin{equation}\label{eqn:poly_chara_diff1}
    \sum_{\dindex=1}^{\mixnumber-1}
       \difference{\cconsttwo_{\dindex}^{(1)}}{1}{\scf_{\dindex}(\cocomp_1+\cconst_{\dindex}\cocomp_2)}
      =\difference{\cconstfour_1}{1}{\cfunc_1(\cocomp_1)}+
       \difference{\frac{-\cconstfour_1}{\cconst_{\mixnumber}}}{1}{\cfunc_2(\cocomp_2)},
  \end{equation}
  where $\difference{}{}{\cdot}$ is the general difference operator defined by
  \begin{equation}
  \begin{split}
    \difference{\cconstthree}{1}{f(\cocomp)} = & f(\cocomp+\cconstthree)-f(\cocomp)\\
    \text{and}\\
    \difference{\cconstthree_0,\ldots,\cconstthree_\altsounumber}{\altsounumber+1}{f(\cocomp)} 
    = &\difference{\cconstthree_0,\ldots,\cconstthree_{\altsounumber-1}}{\altsounumber}
        {f(\cocomp+\cconstthree_\altsounumber)-f(\cocomp)}
  \end{split}
  \end{equation}
  for any constants $\cconstthree_{\dindex}\in\mathbb{C}$. Equation~(\ref{eqn:poly_chara_diff1})
  is of the same form as (\ref{eqn:poly_characterization}) except the number of the
  terms in the sum is lower.
  Let $\cconsttwo_{\dindex}^{(2)}=(1-\frac{\cconst_{\dindex}}{\cconst_{\mixnumber-1}})\cconstfour_2$.
  Again by substituting and subtracting, we obtain from (\ref{eqn:poly_chara_diff1}) the equation
  \begin{equation}\label{eqn:poly_chara_diff2}
    \sum_{\dindex=1}^{\mixnumber-2}
      \difference{\cconsttwo_{\dindex}^{(1)},\cconsttwo_{\dindex}^{(2)}}{2}
      {\scf_{\dindex}(\cocomp_1+\cconst_{\dindex}\cocomp_2)}
      =\difference{\cconstfour_1,\cconstfour_2}{2}{\cfunc_1(\cocomp_1)}+
       \difference{\frac{-\cconstfour_1}{\cconst_{\mixnumber}},
                   \frac{-\cconstfour_2}{\cconst_{\mixnumber-1}}}{2}{\cfunc_2(\cocomp_2)}.
  \end{equation}
  Continuing the process, we end up with the equation
  \begin{equation}\label{eqn:poly_chara_difflast}
  \begin{split}
      \difference{\cconsttwo_{1}^{(1)},\ldots,\cconsttwo_{1}^{(\mixnumber-1)}}{\mixnumber-1}
      {\scf_{1}(\cocomp_1+\cconst_{1}\cocomp_2)} & \\
    =  \difference{\cconstfour_1,\ldots,\cconstfour_{\mixnumber-1}}{\mixnumber-1}{\cfunc_1(\cocomp_1)}
    + & \difference{\frac{-\cconstfour_1}{\cconst_{\mixnumber}},\ldots,
                   \frac{-\cconstfour_{\mixnumber-1}}{\cconst_{2}}}{p-1}{\cfunc_2(\cocomp_2)}.
  \end{split}
  \end{equation}
  This is the generalized Cauchy's equation for complex variables \cite{Aczel:1989} showing that
  $\difference{\cconsttwo_{1}^{(1)},\ldots,\cconsttwo_{1}^{(\mixnumber-1)}}{\mixnumber-1}
    {\scf_{1}(\cocomp)}=\cconstthree\cocomp+\cconstfour\cconj{\cocomp}$ for some constants
    $\cconstthree,\cconstfour\in\mathbb{C}$. Since coefficients $\cconstfour_{\dindex}$
    are arbitrary in the neighborhood
    of zero, and by continuity, the difference operator structure \cite{Aczel:1966} shows that 
    $\scf_{1}(\cocomp)$ is a wide sense polynomial in $(\cocomp,\cconj{\cocomp})$ of
    degree not exceeding $\mixnumber$. By renumbering, the same is obtained for
    $\scf_{\dindex}(\cocomp)$, $\dindex=1,\ldots,\mixnumber$, and thus also
    for $\cfunc_1(\cocomp)$ and $\cfunc_2(\cocomp)$.
\end{proof}

\begin{proofwl}{Proof of Theorem~\ref{thm:complexDS}}
  The joint c.f. of $(\mixturecomp_1,\mixturecomp_2)^\trans$ is given as
  \begin{equation}\label{eqn:cf_eqn1}
  \begin{split}
    ´& \cf_{\mixturecomp_1,\mixturecomp_2} \bigl(\cocomp_1,\cocomp_2\bigr)\\
    = &\E{\mixturecomp_1,\mixturecomp_2}{\exp\bigl(\jmath
         \real{\inner{(\cocomp_1,\cocomp_2)^\trans}{(\mixturecomp_1,\mixturecomp_2)^\trans}}\bigr)}\\
    = &\E{\mixturecomp_1,\mixturecomp_2}{\exp\bigl(\jmath
         \real{\inner{(\cocomp_1,\cocomp_2)^\trans}{\sum_{\dindex=1}^\altsounumber
        (\mixingel_\dindex\sourcecomp_\dindex,
         \altmixingel_\dindex\sourcecomp_\dindex)^\trans}}\bigr)}\\
    = &\E{\mixturecomp_1,\mixturecomp_2}{\exp\bigl(\jmath
         \sum_{\dindex=1}^\altsounumber\real{
         (\mixingel_\dindex\cocomp_1+\altmixingel_\dindex\cocomp_2)\sourcecomp_\dindex}\bigr)}\\
    = &\prod_{\dindex=1}^\altsounumber\E{\sourcecomp_\dindex}{\exp\bigl(\jmath
         \real{(\mixingel_\dindex\cocomp_1+\altmixingel_\dindex\cocomp_2)\sourcecomp_\dindex}\bigr)}\\
    = &\prod_{\dindex=1}^\altsounumber\cf_{\sourcecomp_\dindex}
     (\mixingel_\dindex\cocomp_1+\altmixingel_\dindex\cocomp_2),
  \end{split}
  \end{equation}
  $\cocomp_1,\cocomp_2\in\mathbb{C}$,
  by independence of r.v.s $\sourcecomp_\dindex$, $\dindex=1,\ldots,\altsounumber$. On the
  other hand, by independence of $\mixturecomp_1$ and $\mixturecomp_2$, we have
  \begin{equation}\label{eqn:cf_eqn2}
  \begin{split}
    \cf_{\mixturecomp_1,\mixturecomp_2}\bigl(\cocomp_1,\cocomp_2\bigr)
    = &\cf_{\mixturecomp_1}(\cocomp_1)\cf_{\mixturecomp_2}(\cocomp_2)\\
    = &\prod_{\dindex=1}^\altsounumber\cf_{\sourcecomp_\dindex}(\mixingel_\dindex\cocomp_1)
     \prod_{\dindex=1}^\altsounumber\cf_{\sourcecomp_\dindex}(\altmixingel_\dindex\cocomp_2).
  \end{split}
  \end{equation}
  Thus by combining equations (\ref{eqn:cf_eqn1}) and (\ref{eqn:cf_eqn2}), we get
  \begin{equation}\label{eqn:cf_eqn}
    \prod_{\dindex=1}^\altsounumber\cf_{\sourcecomp_\dindex}
     (\mixingel_\dindex\cocomp_1+\altmixingel_\dindex\cocomp_2)
    =\prod_{\dindex=1}^\altsounumber\cf_{\sourcecomp_\dindex}(\mixingel_\dindex\cocomp_1)
     \prod_{\dindex=1}^\altsounumber\cf_{\sourcecomp_\dindex}(\altmixingel_\dindex\cocomp_2).
  \end{equation}
  As always, there exists a neighborhood of zero such that all c.f.s in Eq.~(\ref{eqn:cf_eqn})
  are nonzero. Let $\altsourcecomp_\dindex=\cconj{\mixingel_\dindex}\sourcecomp_\dindex$ and 
  $\cconst_\dindex=\altmixingel_\dindex/\mixingel_\dindex$ for $\mixingel_\dindex\neq0$,
  and $\cconst_\dindex=\altmixingel_\dindex$ for $\mixingel_\dindex=0$.
  Then, by Eq.~(\ref{eqn:htrans}), we can 
  rewrite Eq.~(\ref{eqn:cf_eqn}) for some positive $\varepsilon>|\cocomp_1|,|\cocomp_2|$
  by setting $\scf_{\dindex}=\log\cf_{\altsourcecomp_\dindex}$ as
  \begin{equation}\label{eqn:scf_eqn}
    \sum_{\dindex=1}^\dindextwo\scf_{\dindex}
     (\cocomp_1+\cconst_\dindex\cocomp_2)
    =\sum_{\dindex=1}^\dindextwo\scf_{\dindex}(\cocomp_1)
     +\sum_{\dindex=1}^\dindextwo\scf_{\dindex}(\cconst_\dindex\cocomp_2),
  \end{equation}
  where it is assumed without loss of generality that 
  $\dindextwo$ first r.v.s $\altsourcecomp_\dindex$, $\dindex=1,\ldots,\dindextwo$,
  are such that $\mixingel_\dindex\altmixingel_\dindex\neq 0$, and therefore
  components $\scf_{\dindex}$, $\dindex>\dindextwo$, cancel out. By combining 
  functions $\scf_{\dindex}$ with the equal arguments to a single function $\tilde{\scf}$
  and renumbering,  Eq.~(\ref{eqn:scf_eqn}) may be rewritten as
  \begin{equation}
    \sum_{\dindex=1}^\dindexthree\tilde{\scf}_{\dindex}
     (\cocomp_1+\cconst_\dindex\cocomp_2)
    =\sum_{\dindex=1}^\dindextwo\scf_{\dindex}(\cocomp_1)
     +\sum_{\dindex=1}^\dindexthree\tilde{\scf}_{\dindex}(\cconst_\dindex\cocomp_2)
  \end{equation}
  such that numbers $\cconst_\dindex$, $\dindex=1,\ldots,\dindexthree\leq\dindextwo$, are distinct.
  Therefore, $\sum_{\dindex=1}^\dindextwo\scf_{\dindex}(\cocomp_1)$ is a
  wide sense polynomial by Lemma~\ref{thm:complexFE}. By Theorem~\ref{thm:marcinkiewicz},
  the r.v. $\sum_{\dindex=1}^{\dindextwo}\altsourcecomp_\dindex$ is complex
  normal. Thus by Theorem~\ref{thm:cramer} each r.v. $\altsourcecomp_\dindex$,
  and hence each r.v. $\sourcecomp_\dindex$, $\dindex=1,\ldots,\dindextwo$, is complex normal.
\end{proofwl}

\section{Additional characterization lemmas}\label{sec:prooflemmas}

\begin{lemma}\label{thm:nonorthogonal}
  Let $\mixingcol_1,\ldots,\mixingcol_\sounumber$ be given nonzero vectors of
  an inner product space. Then there exist a vector $\altmixingcol$, which
  is not orthogonal to any of the given vectors.
\end{lemma}
\begin{proof}
   Suppose $\altmixingcol$ is not orthogonal to any
   $\mixingcol_\dindextwo$, $\dindextwo=1,\ldots,\dindex-1$, but 
   is orthogonal to $\mixingcol_\dindex$. 
   Then a scalar $\cconst\in\mathbb{C}$ can be chosen such that 
   $\inner{\altmixingcol}{\mixingcol_\dindextwo}\neq-
   \cconst\inner{\mixingcol_\dindex}{\mixingcol_\dindextwo}$
   for all $\dindextwo\leq \dindex$. Now the vector 
   $\hat{\altmixingcol}=\altmixingcol+\cconst\mixingcol_\dindex$ is not
   orthogonal to any $\mixingcol_\dindextwo$, $\dindextwo\leq \dindex$. 

   Since $\mixingcol_1$ is nonzero,
   $\altmixingcol_1=\mixingcol_1$ is not orthogonal to $\mixingcol_1$. 
   Choose $\altmixingcol_2=\altmixingcol_1+\cconst_2\mixingcol_2$, 
   where $\cconst_2$ is a scalar as above
   if $\altmixingcol_1$ is orthogonal to $\mixingcol_2$, and  $\cconst_2=0$ otherwise. 
   By iterating the procedure $\sounumber-1$ times,
   it is seen that $\altmixingcol_\sounumber$ is a required type of vector.
\end{proof}

\begin{lemma}\label{thm:twodimmatrices}
  Let $\mixingcol_1,\ldots,\mixingcol_\sounumber$ be given 
  $\mixnumber$-dimensional nonzero complex vectors such that
  $\mixingcol_1$ is not collinear with any $\mixingcol_\dindex$, $\dindex\neq 1$.
  Then there exists a $2\times \mixnumber$ matrix $\compmatrix$ such that
  $\compmatrix\mixingcol_1$ is not collinear with any $\compmatrix\mixingcol_\dindex$, $\dindex\neq 1$. 
\end{lemma}
\begin{proof}
  Denote $\mixingcol_\dindex=(\mixingel_{\dindex1},\ldots,\mixingel_{\dindex\mixnumber})^\trans$, 
  $\dindex=1,\ldots,\sounumber$. 
  Without loss of generality we assume that the coefficients
  $\mixingel_{\dindex1}$, $\dindex=1,\ldots,\sounumber$, are either zero or one. 
  Furthermore, we may 
  take $\mixingel_{11}=1$ by permutating the original indices.
  
  Suppose $\mixingcol_1$ is not collinear with $\mixingcol_\dindex$, i.e.,
  $\mixingcol_1\neq\mixingcol_\dindex$, for any $\dindex\neq 1$. Define
  \begin{equation}
     \compmatrix=\begin{pmatrix}
                   1 &        0 & \cdots & 0\\
                   \altmixingel_1 & \altmixingel_2 & \cdots & \altmixingel_\mixnumber
                \end{pmatrix},
  \end{equation}
  where $\altmixingcol=(\altmixingel_1,\ldots,\altmixingel_\mixnumber)^\trans$ is a vector
  such that
  \begin{equation}
   \inner{\altmixingcol}{\cconj{(\mixingcol_1-\mixingcol_\dindex)}}\neq 0,\quad \dindex=2\ldots,\sounumber.
  \end{equation}
  By Lemma~\ref{thm:nonorthogonal} such a vector $\altmixingcol$ exists.
  Now vectors $\compmatrix\mixingcol_\dindex$ are again such that the first component is either
  zero or one.  Thus $\compmatrix\mixingcol_1$ can be collinear with another
  vector  $\compmatrix\mixingcol_\dindex$  only if $\mixingel_{\dindex1}=1$.
  But then the difference
  \begin{equation}
   \compmatrix\mixingcol_1-\compmatrix\mixingcol_\dindex 
      =\begin{pmatrix}
        1\\ 
        \altmixingcol^\trans\mixingcol_1
       \end{pmatrix}
      -\begin{pmatrix}
        1\\ 
        \altmixingcol^T\mixingcol_\dindex
       \end{pmatrix}
      =\begin{pmatrix}
        0\\ 
        \inner{\altmixingcol}{\cconj{(\mixingcol_1-\mixingcol_\dindex)}}
       \end{pmatrix}
  \end{equation}
  is not zero by construction. Thus $\compmatrix\mixingcol_1$ is not collinear with
  any $\compmatrix\mixingcol_\dindex$, $\dindex\neq 1$, and $\compmatrix$ is a required type
  of matrix.
\end{proof}

\begin{lemma}\label{thm:twodim_characterization}
  Let $(\mixing,\source)$ and $(\altmixing,\altsource)$
  be two reduced representations of a $2$-dimensional complex r.vc. $\mixture$,
  where $\mixing$ and $\altmixing$ are constant complex matrices of dimensions
  $2\times\sounumber$ and $2\times\altsounumber$ respectively, and
  $\source=(\sourcecomp_1,\ldots,\sourcecomp_\sounumber)^\trans$ and 
  $\altsource=(\altsourcecomp_1,\ldots,\altsourcecomp_\altsounumber)^\trans$ are 
  complex r.vc.s with independent components. Then the following properties hold.
  \begin{enumerate}[(i)]
    \item If the $\dindex$th column of $\mixing$ is not collinear
          with any column of $\altmixing$,
          then the r.v. $\sourcecomp_\dindex$ is complex normal. \label{thm:tc_part1}
    \item If the $\dindex$th column of $\mixing$ is collinear with
          the $\dindextwo$th column of $\altmixing$,
          then the logarithms of the c.f.s of
          $\sourcecomp_\dindex$ and $\altsourcecomp_\dindextwo$ differ
          by a wide sense polynomial in a neighborhood of the origin.\label{thm:tc_part2}
  \end{enumerate}
\end{lemma}
\begin{proof}
\begin{enumerate}[(i)]
  \item Without loss of generality we assume that matrices $\mixing$
        and $\altmixing$ are scaled such that the first rows
        consist only of zeros and ones. This amounts only to the scale
        of r.v.s $\sourcecomp_\dindextwo$ and r.v.s $\altsourcecomp_\dindextwo$. 
        Furthermore, since the components of $\mixture$ can be interchanged if necessary,
        the first entry of the $\dindex$th column of $\mixing$ can be taken to be one.
        
        As always, there exists a neighborhood $\varepsilon>0$
        of zero such that all c.f.s are nonzero, and
        the logarithms of c.f.s are well-defined. Therefore for  
        $\compvector=(\cocomp_1,\cocomp_2)^\trans\in\mathbb{C}^2$,
        $|\cocomp_1|<\varepsilon$, $|\cocomp_2|<\varepsilon$, 
        we have using the properties (\ref{eqn:htrans}) and (\ref{eqn:indi}) that
        \begin{align}
           \log\cf_{\mixture}(\compvector) 
              = &\log\cf_{\source}(\mixing^\htrans\compvector)
                =\log\cf_{\altsource}(\altmixing^\htrans\compvector)\notag\\
              = &\sum_{\dindextwo=1}^\sounumber
                 \log\cf_{\sourcecomp_\dindextwo}(\cconj{\mixingel_{1\dindextwo}}\cocomp_1
                 +\cconj{\mixingel_{2\dindextwo}}\cocomp_2) \label{eqn:twodimeq1}\\ 
              = &\sum_{\dindextwo=1}^\altsounumber
                 \log\cf_{\altsourcecomp_\dindextwo}(\cconj{\altmixingel_{1\dindextwo}}\cocomp_1
                 +\cconj{\altmixingel_{2\dindextwo}}\cocomp_2) \label{eqn:twodimeq2},
        \end{align}
        where $\mixing=(\mixingel_{\dindexthree\dindextwo})$, 
        $\altmixing=(\altmixingel_{\dindexthree\dindextwo})$.
        Let $\dindexthree$ 
        be the number of different noncollinear columns with nonzero coefficients in
        $\mixing$ and $\altmixing$ other than the $\dindex$th column of $\mixing$.
        Now substituting (\ref{eqn:twodimeq2}) from (\ref{eqn:twodimeq1}), and
        combining the terms with equal nonzero coefficient arguments to functions $h_\dindextwo$,
        and with one zero coefficient to $f$ and $g$, respectively,
        we get an equation of the form
        \begin{equation}
          \log\cf_{\sourcecomp_\dindex}(\cocomp_1+\cconj{\mixingel_{2\dindex}}\cocomp_2)
          +\sum_{\dindextwo=1}^\dindexthree h_\dindextwo(\cocomp_1+\gamma_\dindextwo\cocomp_2)
          =f(\cocomp_1)+g(\cocomp_2)
        \end{equation}
        if $\mixingel_{2\dindex}\neq0$, and of the form
        \begin{equation}
          \sum_{\dindextwo=1}^\dindexthree h_\dindextwo(\cocomp_1+\gamma_\dindextwo\cocomp_2)
                    =\log\cf_{\sourcecomp_\dindex}(\cocomp_1)+g(\cocomp_2)
        \end{equation}
        if $\mixingel_{2\dindex}=0$.
        Numbers $\mixingel_{2\dindex},\gamma_1,\ldots,\gamma_\dindexthree$ are now distinct, and
        then by Lemma~\ref{thm:complexFE}, $\log\cf_{\sourcecomp_\dindex}$ must be a
        wide sense polynomial in $(\cocomp,\cconj{\cocomp})$ of degree not exceeding $\dindexthree$. Thus by  
        Theorem~\ref{thm:marcinkiewicz},
        the r.v. $\sourcecomp_\dindex$ is complex normal.
  \item By definitions of representations, $\dindex$th column of $\mixing$ is collinear
        \emph{only} with the $l$th column of $\altmixing$. Thus one of the $h$'s in the proof
        of part (\ref{thm:tc_part1}) is the difference the logarithms of the c.f.s of
        $\sourcecomp_\dindex$ and $\altsourcecomp_\dindextwo$, and the claim follows from Lemma~\ref{thm:complexFE}.
\end{enumerate}
\end{proof}

\begin{lemma}\label{thm:sum_independence}
  Suppose independent complex r.v.s $\sourcecomp_1$ and $\sourcecomp_2$ are independent of
  complex normal r.v.s $\normal_1$ and $\normal_2$. If
  $\sourcecomp_1+\normal_1$ is independent of  $\sourcecomp_2+\normal_2$, then also
  $\normal_1$ and $\normal_2$ are independent. 
\end{lemma}
\begin{proof}
  Since the r.vc. $(\sourcecomp_1,\sourcecomp_2)^\trans$ is independent of
  the r.vc. $(\normal_1,\normal_2)^\trans$, the joint c.f. can be written as
  \begin{equation}    
  \begin{split} \cf_{\sourcecomp_1+\normal_1,\sourcecomp_2+\normal_2}& 
   \bigl(\cocomp_1,\cocomp_2\bigr)\\
   = & \cf_{\sourcecomp_1,\sourcecomp_2}\bigl(\cocomp_1,\cocomp_2\bigr)
       \cf_{\normal_1,\normal_2}\bigl(\cocomp_1,\cocomp_2\bigr)\\
   = & \cf_{\sourcecomp_1}(\cocomp_1)\cf_{\sourcecomp_2}(\cocomp_2)
       \cf_{\normal_1,\normal_2}\bigl(\cocomp_1,\cocomp_2\bigr).
  \end{split}
  \end{equation}
  On the other hand, using the independence of $\sourcecomp_1+\normal_1$ and $\sourcecomp_2+\normal_2$, 
  we have
  \begin{equation}
  \begin{split}
  \cf_{\sourcecomp_1+\normal_1,\sourcecomp_2+\normal_2}\bigl(\cocomp_1,\cocomp_2\bigr)
    = & \cf_{\sourcecomp_1+\normal_1}(\cocomp_1)\cf_{\sourcecomp_2+\normal_2}(\cocomp_2)\\
    = & \cf_{\sourcecomp_1}(\cocomp_1)\cf_{\normal_1}(\cocomp_1)
      \cf_{\sourcecomp_2}(\cocomp_2)\cf_{\normal_2}(\cocomp_2),
  \end{split}
  \end{equation}
  and therefore
  \begin{equation}\label{eqn:sum_independence}
  \begin{split}
     \cf_{\sourcecomp_1}(\cocomp_1)\cf_{\sourcecomp_2}(\cocomp_2)
      & \cf_{\normal_1,\normal_2}\bigl(\cocomp_1,\cocomp_2\bigr)\\
     = &
     \cf_{\sourcecomp_1}(\cocomp_1)\cf_{\sourcecomp_2}(\cocomp_2)
     \cf_{\normal_1}(\cocomp_1)\cf_{\normal_2}(\cocomp_2).
  \end{split}
  \end{equation}
  Then, in some neighborhood of zero, all c.f.s in (\ref{eqn:sum_independence}) are
  nonzero, and we have
  \begin{equation}
     \cf_{\normal_1,\normal_2}\bigl(\cocomp_1,\cocomp_2\bigr)=
     \cf_{\normal_1}(\cocomp_1)\cf_{\normal_2}(\cocomp_2)
  \end{equation}
  in the neighborhood.  
  By the $\alpha$-decomposition theorem \cite[Theorem 7.4.2]{Linnik:1977},
  the equation if valid for all $\cocomp_1$ and $\cocomp_2$, i.e., 
  $\normal_1$ and $\normal_2$ are independent.
\end{proof}

\begin{lemma}\label{thm:degen_independence}
  If complex r.v.s $\normal$ and $\sourcecomp$ are independent and
  $\normal+\sourcecomp$ is independent of $\normal$, then
  $\normal$ is degenerate (i.e., a constant).
\end{lemma}
\begin{proof}
 By Theorem~\ref{thm:complexDS} the r.v. $\normal$ is complex normal.
 As in the proof of Lemma~\ref{thm:sum_independence}, it follows
 that the equation
 \begin{equation}
     \cf_{\normal}(\cocomp_1+\cocomp_2)=
     \cf_{\normal}(\cocomp_1)\cf_{\normal}(\cocomp_2)
  \end{equation}
  is satisfied in a neighborhood of zero. This is only possible if $\normal$ is a degenerate
  complex normal r.v., i.e., a complex normal r.v. with zero variance.
\end{proof}

\begin{biography}[{\includegraphics[width=1in,height=1.25in,clip,keepaspectratio]{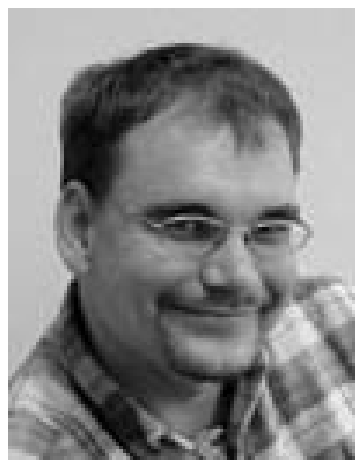}}]{Jan Eriksson}
(M'04) received the M.Sc. degree in mathematics from University of Turku,
Finland, in 2000, and the D.Sc.(Tech) degree (with honors) in signal processing from Helsinki
University of Technology (HUT), Finland, in 2004. He is currently working 
as a postdoctoral researcher of Academy of Finland.

His research interest are in blind signal processing, stochastic modeling, 
digital communication, and  information theory.
\end{biography}

\begin{biography}[{\includegraphics[width=1in,height=1.25in,clip,keepaspectratio]{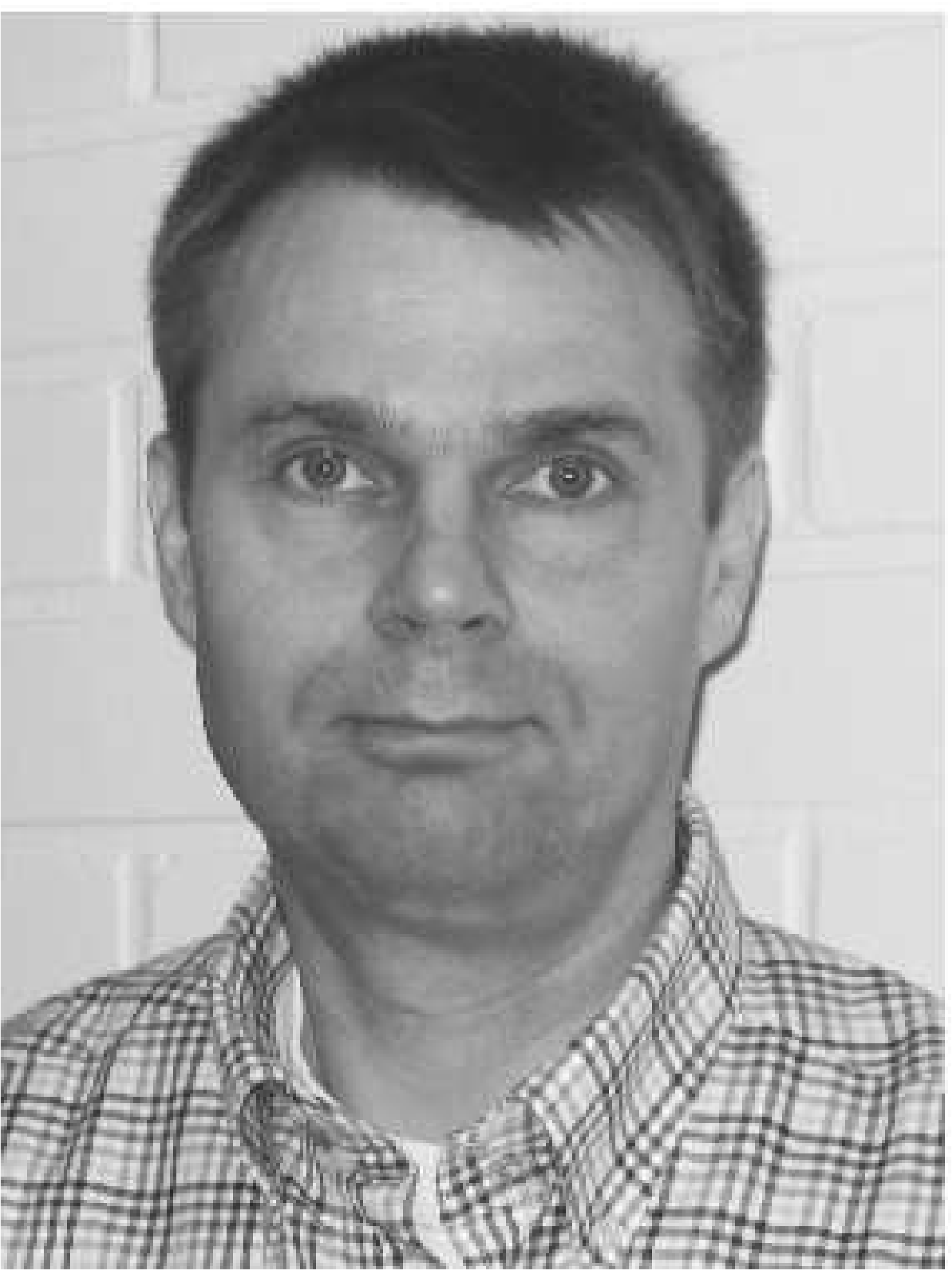}}]{Visa Koivunen}
(Senior Member, IEEE) received his  D.Sc. (Tech) degree with honors  from  the  University  of  Oulu,
Dept.   of Electrical Engineering. From  1992 to 1995 he  was a visiting  researcher at the University of Pennsylvania,
Philadelphia, USA.   Year 1996 he held a faculty  position   at the  Department   of  Electrical Engineering, University
of  Oulu, Finland.  From 1997 to  1999 he was  an Associate  Professor  at the  Signal Processing Labroratory, Tampere
University  of Technology.   Since 1999  he has been  a Professor of Signal Processing  at the Department of  Electrical
and Communications
Engineering, Helsinki  University of Technology (HUT),  Finland. He is
one of the  Principal Investigators in SMARAD Center  of Excellence in
Radio  and  Communications Engineering  nominated  by  the Academy  of
Finland.  Since year 2003 he has been also adjunct professor at the University of Pennsylvania, Philadelphia, USA.
                                                                                                                              
Dr. Koivunen's  research interest include  statistical, communications and sensor array signal processing.  He received
the best paper award (co-authored by C. Ribeiro and A. Richter) from IEEE PIMRC 2005 for his work on MIMO channel
propagation parameter estimation. He has published more than 170 papers in international scientific  conferences  and
journals. He has served as an associate editor for IEEE Signal Processing Letters.
He  is a member  of the editorial board for the Signal Processing journal. He is also a member of the IEEE Signal
Processing for Communication Technical Committee (SPCOM-TC).
\end{biography}

\end{document}